\documentclass[a4paper,11pt]{article}
\pdfoutput=1
\usepackage{jcappub}
\usepackage{slashed}
\usepackage[normalem]{ulem}
\graphicspath{{plots/}}
\title{Heavy Neutral Leptons\\without Prejudice}
\author{Nicolás Bernal,}
\author{Kuldeep Deka,}
\author{and Marta Losada}

\affiliation{New York University Abu Dhabi\\
PO Box 129188, Saadiyat Island, Abu Dhabi, United Arab Emirates}

\emailAdd{nicolas.bernal@nyu.edu}
\emailAdd{kuldeep.deka@nyu.edu}
\emailAdd{marta.losada@nyu.edu}

\abstract{
Heavy Neutral Leptons (HNLs) provide a compelling extension to the Standard Model, addressing the neutrino masses, baryogenesis, and dark matter problems. We perform a model-independent collider study, decoupling the active-sterile mixing angle ($V$) from the Yukawa coupling ($y$), and explore sensitivities at the HL-LHC for prompt and displaced decays. We also consider the possibility of HNLs being long-lived particles decaying in far detectors as FASER. In addition, we study the expected reach at FCC-ee for the prompt and displaced cases. For zero mixing, FCC-ee and HL-LHC sensitivities to $y$ are comparable, with Higgs width measurements imposing the strongest constraints. With non-zero mixing, sensitivities are dominated by $V$, significantly constraining  parameter space. This work highlights the importance of precision Higgs studies and displaced searches in probing HNLs at current and future colliders.}
\begin{document}
\begin{flushright}
\end{flushright}
\maketitle

%%%%%%%%%%%%%%%%%%%%%%%%%%%%%%%%%%%%%
\section{Introduction}
%%%%%%%%%%%%%%%%%%%%%%%%%%%%%%%%%%%%%
The discovery of the Higgs boson in 2012~\cite{ATLAS:2012yve, CMS:2012qbp} provided significant support for the Standard Model (SM), enhancing its alignment with the observed data. Among the many triumphs of the SM are its predictions of gluons~\cite{Barber:1979yr}, the $W^\pm$ and $Z$ bosons' existence and masses~\cite{UA1:1983crd}, and the identification of charm and top quarks~\cite{E598:1974sol, Roy:1989zd, CDF:1994juo}. However, some critical questions remain unresolved, including the origins of neutrino masses, dark matter, matter-antimatter asymmetry, and a few experimental anomalies. To address these gaps, various beyond-SM (BSM) models have been proposed, with precision Higgs physics offering potential avenues for constraint, leveraging both the HL-LHC and future Higgs factories.

Neutrino oscillation experiments have confirmed that neutrinos have mass, raising questions about potential connections to the Higgs mechanism, which is responsible for mass generation of the other SM particles. One promising BSM extension involves the addition of Heavy Neutral Leptons (HNLs), which interact with active SM neutrinos and the Higgs boson through Yukawa couplings. Depending on the model, these sterile HNLs can mix with SM neutrinos post-electroweak symmetry breaking when the Higgs acquires a vacuum expectation value (vev)~\cite{Abazajian:2012ys, Abdullahi:2022jlv}.

Several mechanisms, such as seesaw models and scotogenic approaches, can generate nonzero neutrino masses~\cite{Minkowski:1977sc, Gell-Mann:1979vob, Yanagida:1979as, Mohapatra:1979ia, Glashow:1979nm, Schechter:1980gr, Schechter:1981cv, Foot:1988aq, Ma:1998dn, Choudhury:2020cpm}. Beyond neutrino mass, HNLs can also contribute to the generation of the baryon asymmetry of the universe via leptogenesis~\cite{Fukugita:1986hr, Akhmedov:1998qx, Asaka:2005pn, Davidson:2008bu, Hambye:2016sby, Deka:2021koh} and may even explain the observed dark matter relic abundance~\cite{Dodelson:1993je, Shi:1998km, Abazajian:2001nj, Cline:2020mdt}. In particular, introducing three HNLs into a minimal neutrino SM framework could address all these challenges~\cite{Asaka:2005an, Asaka:2005pn}.

Instead of focusing on a specific model, this study adopts a model-independent approach to experimentally test HNLs. Many searches have probed high-mass HNLs through their production or prompt decays in colliders~\cite{Abreu:1996pa, Aad:2011vj, Chatrchyan:2012fla, Khachatryan:2015gha, Aad:2015xaa, Khachatryan:2016olu, Sirunyan:2018mtv}, see also Refs.~\cite{delAguila:2007qnc, delAguila:2008cj, delAguila:2008hw, Atre:2009rg, BhupalDev:2012zg, Dev:2013wba, Das:2014jxa, Alva:2014gxa, Deppisch:2015qwa, Banerjee:2015gca, Arganda:2015ija, Das:2015toa, Degrande:2016aje, Mitra:2016kov, Das:2017zjc, Das:2017rsu, Ruiz:2017yyf, Cai:2017mow, Accomando:2017qcs, Drewes:2018gkc, Pascoli:2018rsg, Bhaskar:2023xkm}. In certain parameter regions, HNLs may have substantially displaced decays within detectors, creating a unique displaced vertex signature~\cite{Gronau:1984ct}. Recent studies suggest that these signatures could be targeted in LHC searches with associated charged leptons~\cite{Nemevsek:2011hz, Helo:2013esa, Izaguirre:2015pga, Dib:2015oka, Dube:2017jgo, Cottin:2018kmq, Cottin:2018nms, Dib:2018iyr, Nemevsek:2018bbt, Abada:2018sfh, Marcano:2018fto, Abada:2018ulr, Dib:2019ztn}, via Higgs decays~\cite{Pilaftsis:1991ug, Maiezza:2015lza, Gago:2015vma, Accomando:2016rpc, Nemevsek:2016enw, Caputo:2017pit, Deppisch:2018eth, Liu:2018wte}, and even in detectors such as LHCb~\cite{Antusch:2017hhu}. Similar efforts are underway at DUNE~\cite{Adams:2013qkq, Gunther:2023vmz}, IceCube~\cite{Coloma:2017ppo}, proposed lepton colliders~\cite{Blondel:2014bra, Antusch:2016vyf}, and SHiP~\cite{Alekhin:2015byh, Anelli:2015pba}; the latter shows particular promise for HNLs below the $c$-quark mass through meson decays~\cite{Bonivento:2013jag}. ATLAS and CMS have already performed some of these analyzes~\cite{CMS:2015qur, CMS:2018iaf, ATLAS:2019kpx, CMS:2022fut, ATLAS:2022atq}.

Additional far detectors at the LHC, such as FASER~\cite{Feng:2017uoz, FASER:2018eoc}, MoEDAL-MAPP~\cite{Pinfold:2019nqj, Pinfold:2019zwp}, MATHUSLA~\cite{Chou:2016lxi, Curtin:2018mvb, MATHUSLA:2020uve}, ANUBIS~\cite{Bauer:2019vqk}, CODEX-b~\cite{Gligorov:2017nwh}, and FACET~\cite{Cerci:2021nlb}, have been proposed to further study long-lived particles, with FASER and MoEDAL-MAPP currently operational. The case of HNLs as long-lived particles has been a focus in recent studies~\cite{Caputo:2016ojx, Jana:2018rdf, Kling:2018wct, Helo:2018qej, Dercks:2018wum, Deppisch:2019kvs, Frank:2019pgk, Jones-Perez:2019plk, Hirsch:2020klk, Li:2023dbs, Bhattacherjee:2023plj, Deppisch:2023sga, Feng:2024zfe}.

Future colliders could significantly enhance the sensitivity to HNLs and related phenomena compared to current facilities. The Future Circular Collider in its electron-positron mode (FCC-ee)~\cite{Agapov:2022bhm} stands out as a promising candidate, due to its ability to achieve high-statistics data across various operational phases and its exceptionally clean experimental environment. This makes FCC-ee particularly well suited for probing both prompt and displaced decay signatures of HNLs.

To date, HNL searches have mainly focused on signatures from mixing with active neutrinos in the SM. However, the reliance on mixing for the HNL phenomenology is highly dependent on the chosen model. In scenarios where Yukawa couplings are dominant, Yukawa interactions could solely characterize the phenomenology of HNL. Thus, we propose an alternative approach that leverages HNL Yukawa interactions with the Higgs boson along with the mixing contribution from gauge bosons. We keep Yukawa couplings and mixing angles independent to minimize model assumptions. The prospects of the setup at the LHC treating mixing angles as negligible and the role of precision Higgs constraints were already studied in Ref.~\cite{Bernal:2023coo}; here, we include the extension of the study to FCC-ee. The aim of the current study presented here is more general, and the results help us delineate the parameter space where one dominates over the other. 

The paper is organized as follows. Section~\ref{sec:framework} outlines the model and analysis setup. Section~\ref{sec:HNL} describes HNL production and decay along with constraints from various experimental measurements. Sections~\ref{sec:hl-lhc-prompt}, \ref{sec:hl-lhc-dv}, and~\ref{sec:hl-lhc-llp} cover analyses for the LHC of HNLs that decay promptly, with displaced vertices, and in far detectors, respectively. Section~\ref{sec:FCCee} deals with the prompt and displaced analyzes at FCC-ee for two key operational phases, $Z$ pole in Section~\ref{sec:FCCee-zpole}, and $Zh$ production in Section~\ref{sec:FCCee-zh}. Finally, Section~\ref{sec:concl} presents our conclusions. In the appendix, we present, where relevant, the recast of existing constraints on mixing in our setup.

%%%%%%%%%%%%%%%%%%%%%%%%%%%%%%%%%%%%%
\section{Set-up} \label{sec:framework}
%%%%%%%%%%%%%%%%%%%%%%%%%%%%%%%%%%%%%
In this work, we extend the SM by introducing $n$ HNLs, labeled $\widetilde{N}$, which can be Dirac or Majorana particles. The relevant Yukawa interaction is given by 
\begin{equation}
    y_{N \alpha}\, \overline{\widetilde{N}}\, \widetilde{H}^\dagger\, L_\alpha\,,
\end{equation}
where $H$ is the SM Higgs doublet, $\widetilde{H} \equiv i \sigma^2 H^*$ (with $\sigma^2$ as the second Pauli matrix), $L_\alpha$ are the SM lepton doublets for $\alpha = e, \mu, \tau$, and $y_{N \alpha}$ are the Yukawa couplings linking HNLs with SM leptons. The mass of each HNL, $m_N$, is introduced, but is not explicitly written here, since it depends on whether $\widetilde{N}$ is Dirac or Majorana.

After electroweak symmetry breaking, the Higgs field acquires a vev $v_h \simeq 246$~GeV, yielding Dirac-type mass terms for the HNLs, defined as 
\begin{equation}
    (M_D)_{N\alpha} \equiv \frac{y_{N\alpha}\, v_h}{\sqrt{2}}\,.
\end{equation}
Similar terms can also arise if additional scalar fields, which gain vevs, are present in the model. The physical mass eigenstates of the three SM neutrinos $\nu$ and the $n$ HNLs are then obtained by diagonalizing the full mass matrix in a manner dictated by the model. To satisfy the neutrino oscillation data, at least two HNLs ($n \geq 2$) are necessary~\cite{Esteban:2020cvm, deSalas:2020pgw}. Additionally, fitting sub-eV neutrino masses~\cite{KATRIN:2021uub} imposes specific constraints on the model's parameter space. Diagonalizing to the mass basis introduces mixings between HNLs and SM neutrinos, which, if present, allows HNLs to interact with other SM particles, especially the $W^\pm$ and $Z$ bosons. For the type-I seesaw mechanism~\cite{Minkowski:1977sc, Gell-Mann:1979vob, Yanagida:1979as, Mohapatra:1979ia, Glashow:1979nm, Schechter:1980gr, Schechter:1981cv}, the mixing angle scales with $y_{N \alpha}\, v_h/m_N$. However, in inverse seesaw and double-seesaw models~\cite{Dias:2012xp, Fraser:2014yha}, an additional mass scale reduces the mixing contributions. In particular, some models can fine-tune mixing contributions to be precisely zero~\cite{Ma:2014qra, CentellesChulia:2016rms, Bernal:2023coo}. In contrast, in scotogenic models of neutrino mass, it is the Yukawa coupling to the SM Higgs that vanishes~\cite{Ma:2006km, Cai:2017jrq}. It is important to note that while some models connect Yukawa couplings and mixing angles, these relationships should not be universally assumed. This had also been emphasized strongly in our previous study~\cite{Bernal:2023coo}.

To maintain a model-independent approach, in the mass basis, we treat the Yukawa couplings, the mixing angles, and the physical HNL masses as separate free parameters, being agnostic about their possible interdependence. This approach allows for a broader exploration of the HNL parameter space. Such a set-up can be particularly relevant for displaced decay searches and precision Higgs studies.

%%%%%%%%%%%%%%%%%%%%%%%%%%%%%%%%%%%%%
\section{Production and Decays of HNLs} \label{sec:HNL}
%%%%%%%%%%%%%%%%%%%%%%%%%%%%%%%%%%%%%
HNLs with masses in the GeV range can be produced via decays of gauge and Higgs bosons. For the production via Higgs, the partial decay width of the Higgs boson into an HNL and an active neutrino is given by
\begin{equation}
    \Gamma(h \to N \nu) = \frac{y^2}{8 \pi}\, m_h \left[1 - \left(\frac{m_N}{m_h}\right)^2 \right]^2,
\end{equation}
where $m_h \simeq 125$~GeV is the Higgs mass, and $y^2 \equiv y_{Ne}^2 + y_{N\mu}^2 + y_{N\tau}^2$ represents the sum of all Yukawa couplings. Since LHC is insensitive to neutrino flavor, $y^2$ covers all contributions. This decay mode affects the total decay width of the Higgs boson, which is predicted in the SM to be $\Gamma_h \simeq 4.1$~MeV~\cite{LHCHiggsCrossSectionWorkingGroup:2016ypw}. Current measurements at the LHC place the Higgs width at $\Gamma_h = 3.2^{+2.4}_{-1.7}~\text{MeV}$~\cite{ParticleDataGroup:2024cfk}. The high-luminosity LHC (HL-LHC) is projected to significantly improve this precision, achieving an uncertainty of just $5.3\%$~\cite{deBlas:2022aow}. It is expected to reduce further to $1\%$ at FCC-ee~\cite{Freitas:2019bre}. In addition, for the $W$- and $Z$-boson decays, the partial widths are
\begin{align}
    \Gamma(W^\pm \to N l_\alpha^\pm) &\simeq \frac{e^2\, V_{N\alpha}^2}{96 \pi\, s_W^2}\, m_W \left[2 - 3 \left(\frac{m_N}{m_W}\right)^2 + \left(\frac{m_N}{m_W}\right)^6 \right], \\
    \Gamma(Z \to N \nu) &= \frac{e^2\, V^2}{96 \pi\, c_W^2\, s_W^2}\, m_Z \left[2 - 3 \left(\frac{m_N}{m_Z}\right)^2 + \left(\frac{m_N}{m_Z}\right)^6 \right],
\end{align}
where $V^2 \equiv V_{Ne}^2 + V_{N\mu}^2 + V_{N\tau}^2$. The SM total decay widths for the $Z$ and $W$ bosons are $\Gamma_Z = (2.4940 \pm 0.0009)$~GeV and $\Gamma_W = (2.0892 \pm 0.0008)$~GeV, respectively, while their decay into SM neutrinos has partial widths $\Gamma(Z \to \nu \bar{\nu}) = (501 \pm 0.045)$~MeV and $\Gamma(W^\pm \to \nu_\alpha l_\alpha^\pm) = 679 \pm 0.12$~MeV~\cite{ParticleDataGroup:2024cfk}. The measured values are pretty close with $\Gamma_Z = (2.4955 \pm 0.0023)$~GeV, $\Gamma_W = (2.085 \pm 0.042)$~GeV, $\Gamma(Z \to \texttt{inv}) = (500 \pm 1.5)$~MeV and $\Gamma(W^\pm \to \nu_\alpha l_\alpha^\pm) = 679 \pm 0.01$~MeV~\cite{ParticleDataGroup:2024cfk}. 

%%%%%%%%%%%%%%%%%%%%%%%%%%%%%%%%%%%%%%%%%%%%%%%%%%%%%%%%%%
\begin{figure}[t!]
    \def\sepf{0.49}
    \centering
    \includegraphics[width=\sepf\columnwidth]{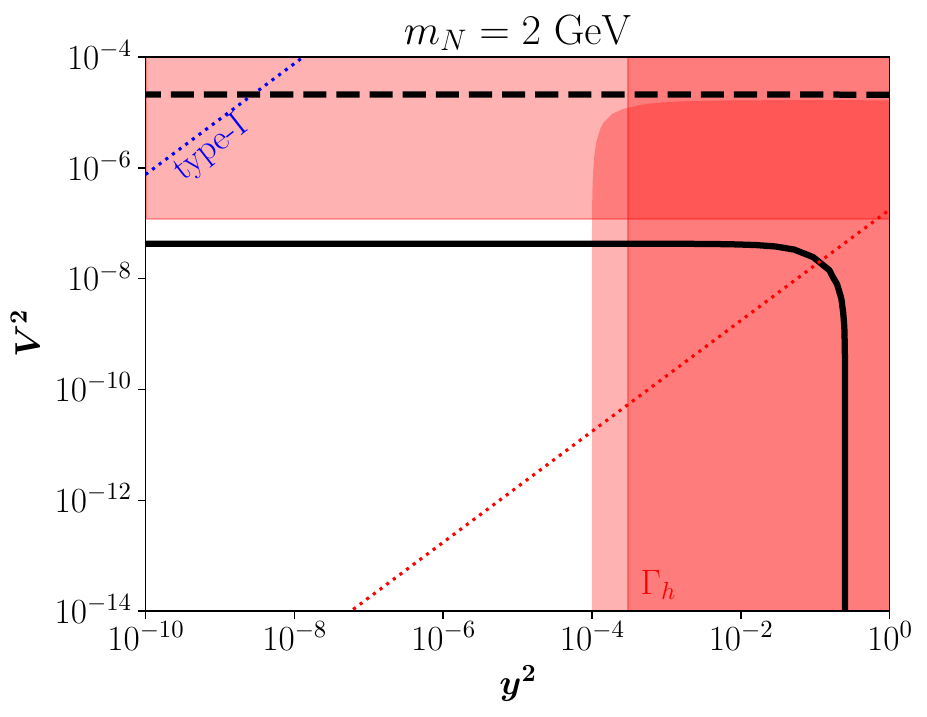}
    \includegraphics[width=\sepf\columnwidth]{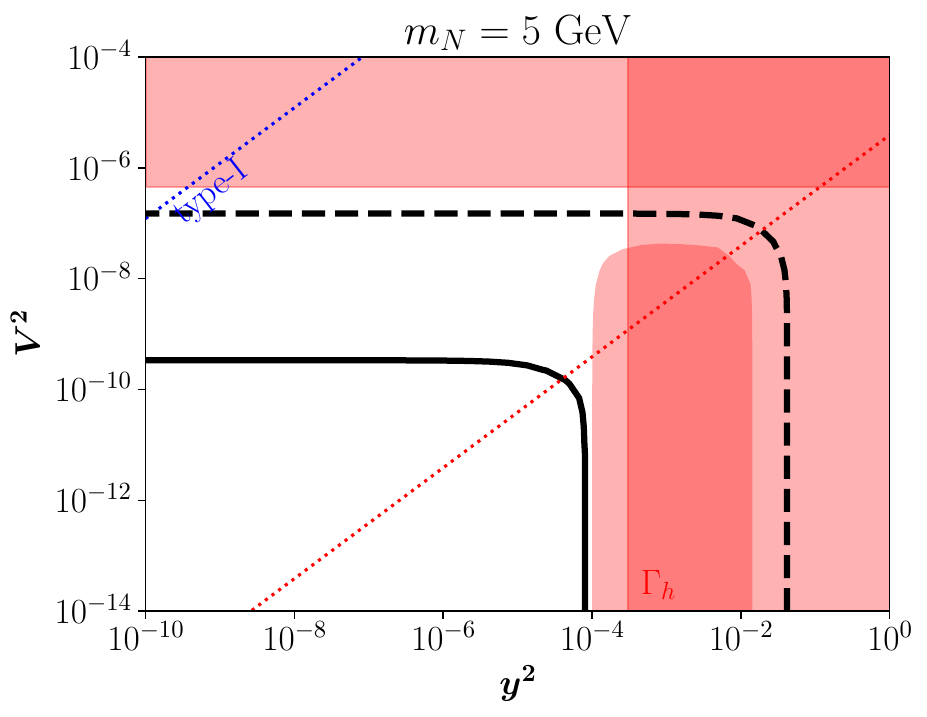}
    \includegraphics[width=\sepf\columnwidth]{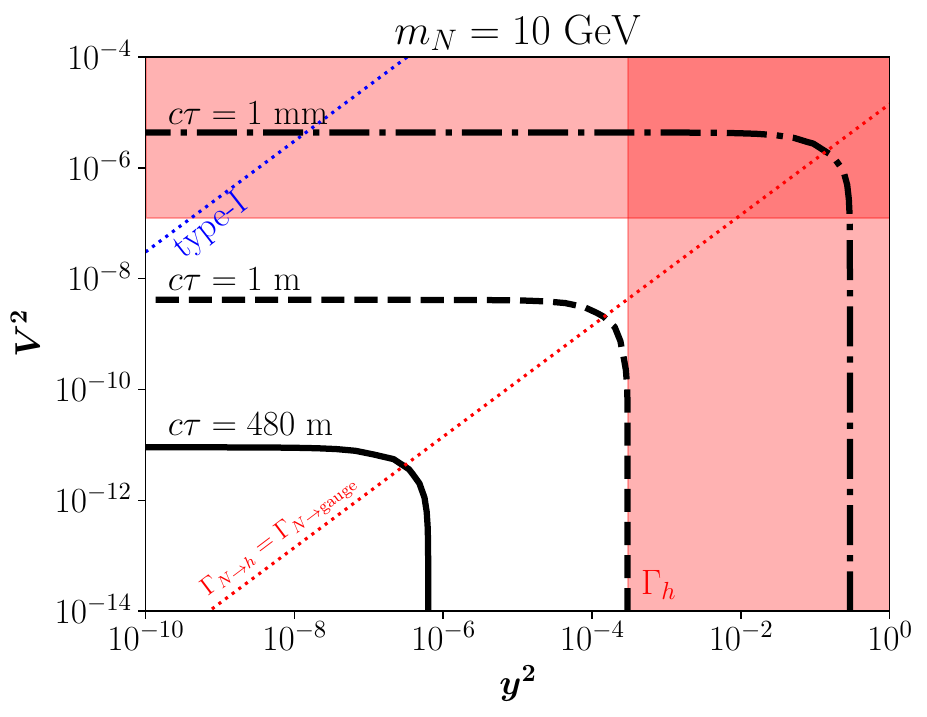}
    \includegraphics[width=\sepf\columnwidth]{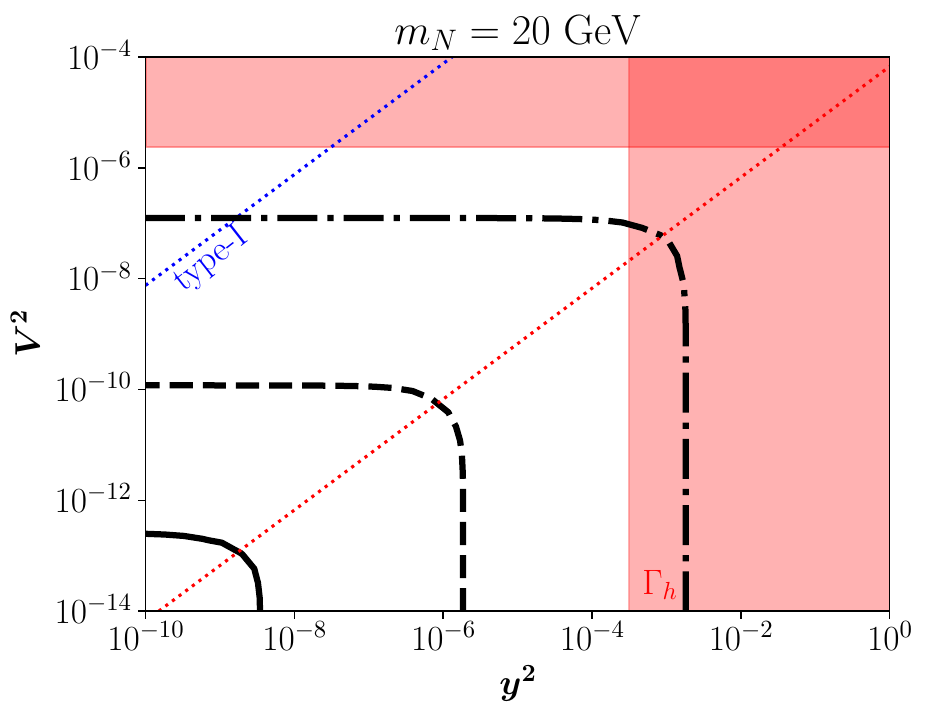}
    \caption{Contour for $c\tau = 1$~mm, 1~m and 480~m, for $m_N =$ 2, 5, 10 and 20~GeV, in the plane $[y^2,V^2]$. The blue dotted line corresponds to the type-I seesaw relation. The red dotted line represents the contour for equal decay width through the Yukawa and mixing. The red bands are in tension with the experimental measurements. The bounds on $V^2$ are taken from the most stringent limits on mixing~\cite{Fernandez-Martinez:2023phj, HNlimits}. The bounds on $y^2$ arise from the total and invisible decay width of the Higgs.}
    \label{fig:ctau-V-y}
\end{figure} 
%%%%%%%%%%%%%%%%%%%%%%%%%%%%%%%%%%%%%%%%%%%%%%%%%%%%%%%%%%
Once produced, HNLs in the GeV ballpark and below decay into three fermions via off-shell gauge or Higgs bosons. We compute these decay widths in \texttt{MADGRAPH5\_aMC@NLO} v2.9.16~\cite{Alwall:2014hca, Frederix:2018nkq} using a modified version of the \texttt{FeynRules}~\cite{Degrande:2011ua, Alloul:2013bka} model \texttt{HeavyN}~\cite{Atre:2009rg, Alva:2014gxa, Degrande:2016aje}. Higgs-mediated decays favor heavy fermions (e.g. $b\bar{b}$, $\tau^+\tau^-$, $c\bar{c}$) for $m_N$ above their thresholds, while gauge boson-mediated decays are universal in this respect. Figure~\ref{fig:ctau-V-y} shows the contours for the proper length $c\, \tau = 1$~mm, 1~m, and 480~m, for HNLs in the plane $[y^2, V^2]$, corresponding to the prompt, displaced and long-lived regimes. Increasing $m_N$ reduces $c\tau \propto 1/m_N^5$, shifting the contours toward lower coupling values. 

The vertical bounds for $y^2$, shown in red, arise from the Higgs total and invisible decay width constraints. Constraints from the invisible branching fraction of the Higgs arise only when the HNL decays through the Yukawa coupling outside the detector~\cite{Bernal:2023coo}. ATLAS and CMS limits on $\text{Br}(h \to \text{inv})$ are $< 10.7\%$ and $< 15\%$, respectively, at 95\% CL~\cite{ATLAS:2023tkt, CMS:2023sdw}, providing stronger constraints than the Higgs total width for $m_N \lesssim 10$~GeV, and are represented by the rectangular vertical red region with rounded edges. These limits are expected to improve to a precision of $1.9 \%$ at HL-LHC and $0.2 \%$ at FCC-ee~\cite{Dawson:2022zbb}. Invisible decays of HNLs~\cite{Atre:2009rg} can also arise through mixing
\begin{equation}
    \Gamma(N \to \nu \nu \nu) = \frac{G_F^2}{96 \pi^3} V^2 m_N^5,
\end{equation}
but it contributes minimally (approximately 6\%)~\cite{Abada:2018sfh}. These decays can be constrained by the invisible width of the $Z$ boson~\cite{ATLAS:2023ynf} to derive limits on $V^2$. However, the most stringent limits on $V^2$ (shown by the horizontal red region) stem from current experimental bounds at colliders~\cite{Fernandez-Martinez:2023phj, HNlimits}. The blue dotted line represents the type-I seesaw relation ($V= y\, v_h/m_N$) as a reference. The red dotted line corresponds to the equal decay width through the mixing and the Yukawa.

%%%%%%%%%%%%%%%%%%%%%%%%%%%%%%%%%%%%%%%%%%%%%%%%%%%%%%%%%%
\section{Prospects at LHC} %\label{sec:prompt}
%%%%%%%%%%%%%%%%%%%%%%%%%%%%%%%%%%%%%%%%%%%%%%%%%%%%%%%%%%
In the following, projections for the HL-LHC will be presented in the case of prompt decays, displaced decays inside the ATLAS or CMS detectors, and long-lived particles.

%%%%%%%%%%%%%%%%%%%%%%%%%%%%%%%%%%%%%%%%%%%%%%%%%%%%%%%%%%
\subsection{Prompt Decays} \label{sec:hl-lhc-prompt}
%%%%%%%%%%%%%%%%%%%%%%%%%%%%%%%%%%%%%%%%%%%%%%%%%%%%%%%%%%
The production of single-Higgs bosons in the LHC is dominated by gluon fusion (ggF) and vector boson fusion (VBF), with corresponding cross sections $\sigma_\text{ggF} \simeq 54.8$~pb and $\sigma_\text{VBF} \simeq 4.26$~pb at $\sqrt{s} = 14$~TeV~\cite{ggF, VBF}. The unique topology of VBF, featuring two forward jets ($j_1, j_2$) in opposite hemispheres, large pseudo-rapidity separation ($\Delta \eta_{j_1 j_2}$), and high invariant mass ($m_{j_1 j_2}$), allows a large suppression of QCD multi-jet backgrounds. We explore single-Higgs boson production via VBF at an integrated luminosity of $\mathcal{L} = 3$~ab$^{-1}$. Subsequently, the Higgs can decay into an HNL and a SM neutrino, with the HNL later decaying into $b\bar{b}$ and another active neutrino via an off-shell Higgs or gauge boson (which can only be $Z$ for the final state of our interest).

The dominant backgrounds include the production of Higgs through VBF and ggF with a later decay $h \to b\bar{b}$, $t\bar{t}$ + jets, and $b\bar{b}$ + jets. The ggF background can mimic the VBF via hard forward jets, despite a distinct topology. The $t\bar{t}$ + jets process is significant due to its large cross section (990~pb at NNLO~\cite{Muselli:2015kba}) and the inherent $b$ jets. Similarly, $b\bar{b}$ + jets, with a cross section of $\sim 10^5$~pb~\cite{ATLAS:2011ac}, can mimic the VBF topology in specific configurations.

To distinguish the signal from the background, we impose the following selection criteria~\cite{Bernal:2023coo}:
\begin{itemize}
    \item \textbf{S1:} Events must have no charged lepton or photon candidates.
    \item \textbf{S2:} At least two $b$-tagged jets and two non-$b$ jets are required.
    \item \textbf{S3:} Leading non-$b$ jets must satisfy $p_{T_{j_1}} > 60$~GeV, $p_{T_{j_2}} > 40$~GeV, and $H_T > 140$~GeV for all non-$b$ jets.
    \item \textbf{S4:} The two leading non-$b$ jets must meet VBF criteria: $\eta_{j_1} \times \eta_{j_2} < 0$, $\Delta \eta_{j_1 j_2} > 3.5$, $m_{j_1 j_2} > 500$~GeV, and $\Delta \phi_{j_1 j_2} < 2.5$.
    \item \textbf{S5:} Additional non-$b$ jets ($j_i$) are tested using the ratio:
    \begin{equation}
        m^\text{rel}_{j_i} \equiv \frac{\text{min}(m_{j_1 j_i},\, m_{j_2 j_i})}{m_{j_1 j_2}}\,,
    \end{equation}
    where smaller values of $m^\text{rel}_{j_i}$ ($m^\text{rel}_{j_i} < 0.08$) indicate compatibility with final-state radiation, helping the rejection of QCD multijet events.
    \item \textbf{S6:} Missing transverse energy $\slashed{E}_T$ must exceed 50~GeV.
    \item \textbf{S7:} The invariant mass of the $b$ jets must satisfy $0.2\, m_N \leq m_{b\bar{b}} \leq 1.6\, m_N$, with $\Delta R_{b \bar{b}} \leq 2.5$.
\end{itemize}

%%%%%%%%%%%%%%%%%%%%%%%%%%%%%%%
\begin{figure}[t!]
    \def\sepf{0.49}
    \centering
    \includegraphics[width=\sepf\columnwidth]{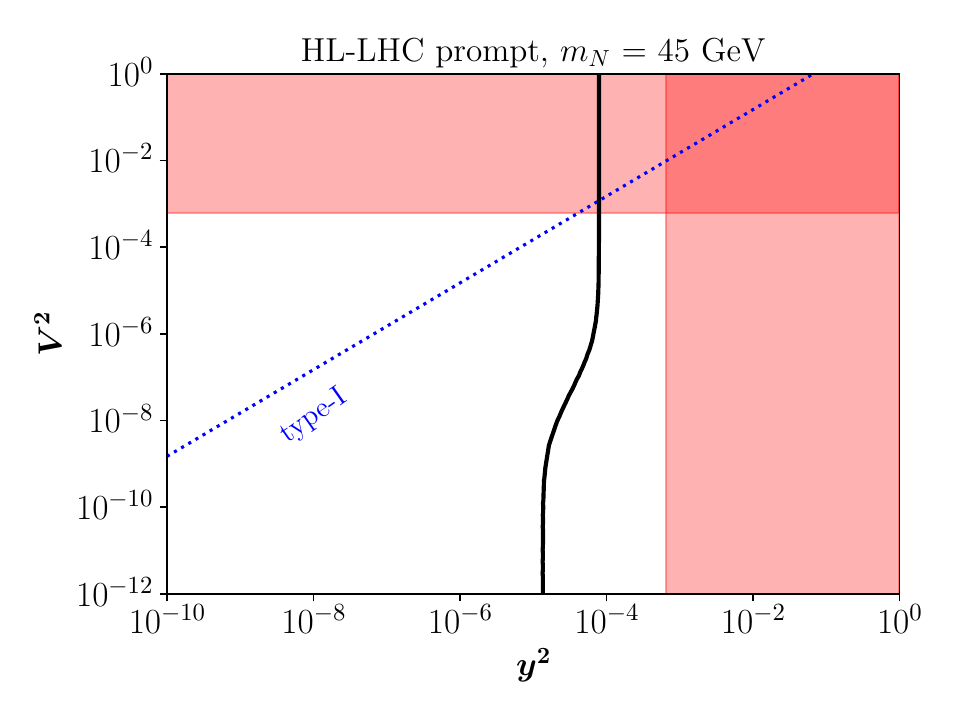}
    \includegraphics[width=\sepf\columnwidth]{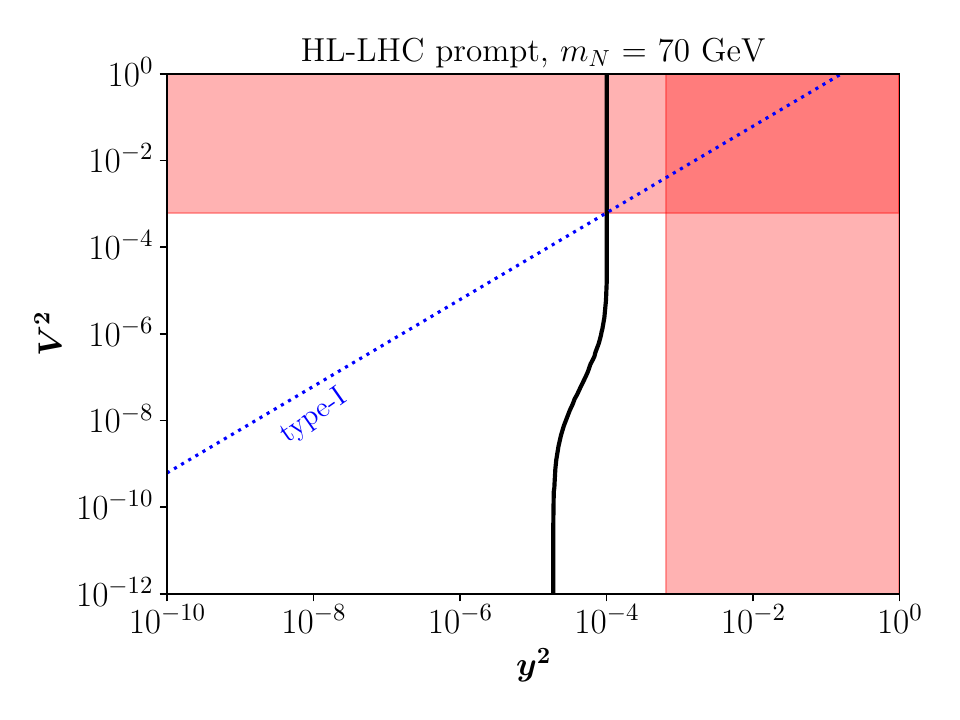}
    \caption{Discovery reach at $3~\sigma$ CL for HL-LHC with VBF Higgs with both Yukawa coupling and mixing angle HNL of mass 45~GeV (left) and 70 GeV (right). The regions to the right of the $3\sigma$ contours will be probed at HL-LHC. The horizontal and vertical red bands arise from the current most-stringent limit on mixing and Higgs total decay width constraint on the Yukawa respectively.}
    \label{fig:HLLHC-prompt}
\end{figure} 
%%%%%%%%%%%%%%%%%%%%%%%%%%%%%%%
Figure~\ref{fig:HLLHC-prompt} presents the discovery reach for the HL-LHC, for two benchmark values of $m_N$, 45~GeV (left) and 70~GeV (right), and the selection cuts previously mentioned. HL-LHC will be able to probe the regions to the right of the $3~\sigma$ contours. Since HNLs are produced via the Higgs boson, achieving a $3\sigma$ significance requires a minimum $y^2$, resulting in the near-vertical contour. As $V^2$ increases, mixing-mediated decays become more efficient, leading to a reduction of the branching fraction of the Higgs into a couple of $b$ quarks, and therefore to a change in the slope of the contour. The red bands show the parameter regions that are in tension with the current experimental limits on mixing and the Higgs total decay width on the Yukawa.

%%%%%%%%%%%%%%%%%%%%%%%%%%%%%%%%%%%%%%%%%%%%%%%%%%%%%%%%%%
\subsection{Displaced Vertices} \label{sec:hl-lhc-dv}
\begin{figure}[t!]
    \def\sepf{0.49}
    \centering
    \includegraphics[width=\sepf\columnwidth]{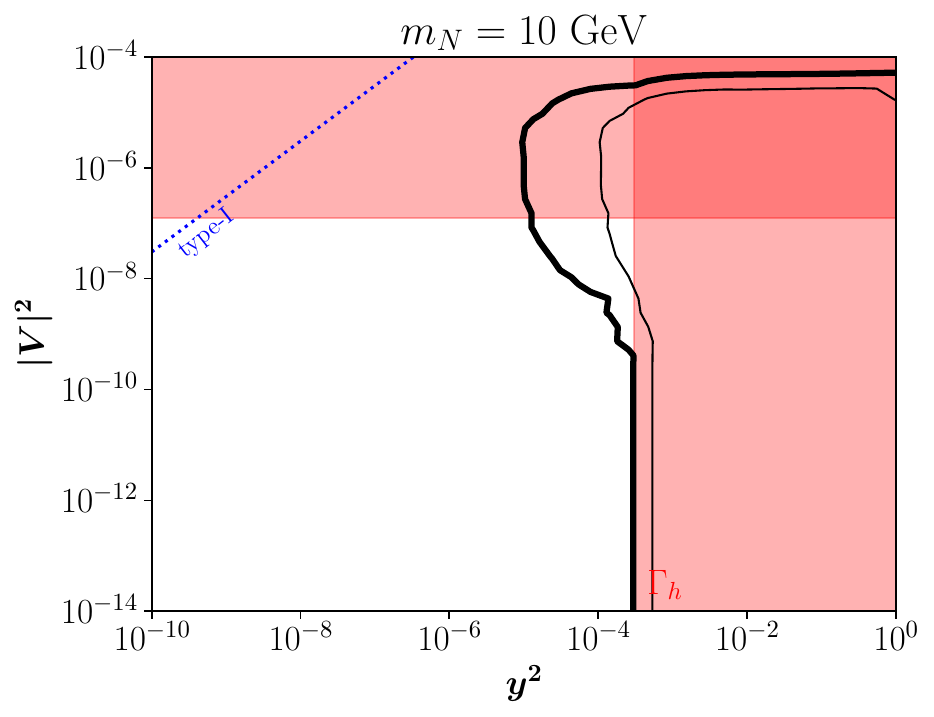}
    \includegraphics[width=\sepf\columnwidth]{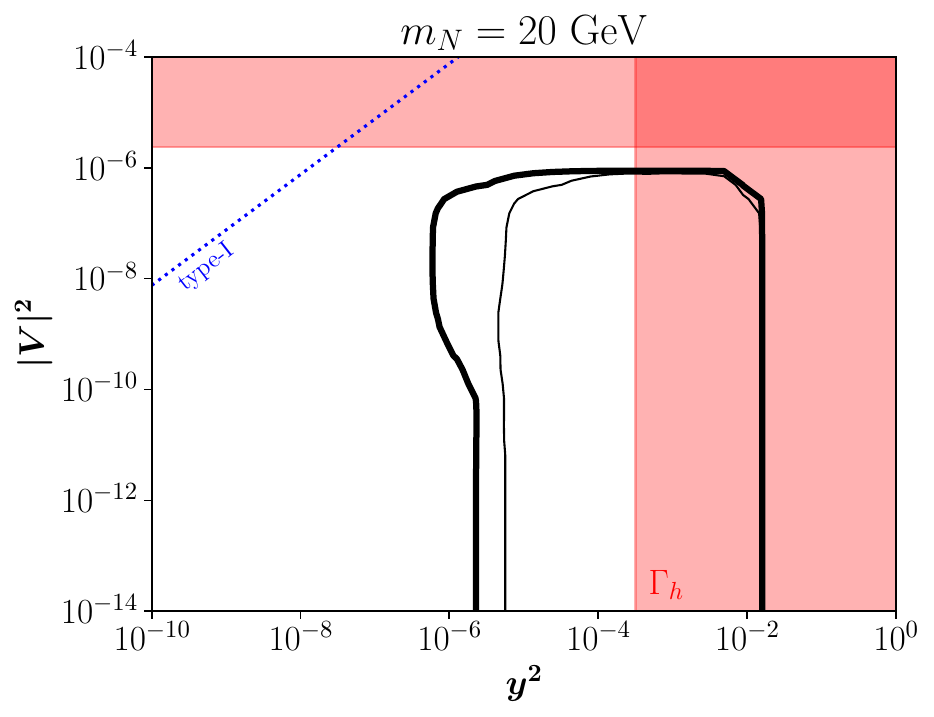}
    \caption{Sensitivity reach of searches for a displaced decay, for $\sqrt{s} = 14$~TeV and $\mathcal{L} = 3$~ab$^{-1}$ are shown by regions inside the contours. The thick line corresponds to the assumption of zero background events, whereas for the thin line 65 events were assumed; see the text for further details. The blue dotted line corresponds to the type-I seesaw relation. The red bands are in tension with the experimental measurements.}
    \label{fig:DV}
\end{figure}

%%%%%%%%%%%%%%%%%%%%%%%%%%%%%%%%%%%%%%%%%%%%%%%%%%%%%%%%%%
This section addresses the scenario in which HNLs decay within the inner tracker of the ATLAS or CMS detectors, resulting in a displaced vertex. As before, we concentrate on Higgs production via VBF to identify the distinct signature of two prominent forward jets. However, ggF events can also significantly meet the same topology criteria, and thus they have been incorporated into the analysis. The selection criteria for the events are as follows:
\begin{itemize}
    \item Same cuts $S_1$, $S_3$ and $S_4$ for the VBF hard forward jets, as described in Section~\ref{sec:hl-lhc-prompt}.
    \item Two extra jets with a displacement of 1~mm $\leq d_{xy} \leq 1$~m and $d_z \leq 300$~mm.
\end{itemize}
Initially, we consider a scenario without background, utilizing a Poisson distribution. We then focus on the parameter space where there are more than 3.09 anticipated signal events at 95\% CL~\cite{ParticleDataGroup:2024cfk}. Figure~\ref{fig:DV} shows the region within a thick solid black line, the sensitivity reach of the searches for a displaced decay for $\sqrt{s} = 14$~TeV and $\mathcal{L} = 3$~ab$^{-1}$, in the plane $[y^2,\, V^2]$, for $m_N=10$~GeV (left) and $m_N=20$~GeV (right). These regions show the same trend as already pointed out in Fig.~\ref{fig:ctau-V-y}, but smaller values of $y^2$ are not allowed due to the insufficient number of HNLs produced.
%%%%%%%%%%%%%%%%%%%%%%%%%%%%%%%%%%%%%%%%%%%%%%%%%%%
%%%%%%%%%%%%%%%%%%%%%%%%%%%%%%%%%%%%%%%%%% 

The hypothesis of having no background previously considered might be overly optimistic, as background events can stem from incorrectly identified displaced events due to detector resolution issues or random track intersections. To get a general idea of the potential impact of including a background on sensitivity, we refer to the discussions in Refs.~\cite{Deppisch:2018eth, Abada:2018sfh, Marcano:2018fto, Abada:2018ulr} and consider the maximum possible background, consistent with the lack of observed background events in the ATLAS analysis found in Refs.~\cite{ATLAS:2022zhj, ATLAS:2023oti}. This corresponds to a maximum of 3 background events for $\mathcal{L} = 139$~fb$^{-1}$, which scales to 65 background events for a luminosity of 3~ab$^{-1}$. This pessimistic case is illustrated in Fig.~\ref{fig:DV} with a thin black line.

%%%%%%%%%%%%%%%%%%%%%%%%%%%%%%%%%%%%%%%%%%%%%%%%%%%%%%%%%%
\subsection{Long-lived Particles} \label{sec:hl-lhc-llp}
%%%%%%%%%%%%%%%%%%%%%%%%%%%%%%%%%%%%%%%%%%%%%%%%%%%%%%%%%%
The FASER detector is located 480 meters downstream of the proton-proton interaction point leveraged by the ATLAS experiment~\cite{Feng:2017uoz, FASER:2018eoc}. Designed to detect new particles within a cylindrical region of radius 10 cm and length 1.5 m, it will initially function at an integrated luminosity of 150 fb$^{-1}$. In contrast, the second phase (FASER-2~\cite{FASER:2018eoc}) at the HL-LHC will have a length of 10~m, a radius of 1~m, and an integrated luminosity of 3 ab$^{-1}$. Our focus here is on this second phase.

Detection of long-lived HNLs can be performed using a detector such as FASER~\cite{Feng:2017uoz}, which is adept at detecting particles decaying approximately 480~m from ATLAS interaction point. This considerable separation ensures minimal background noise, often deemed negligible. Consequently, the experiment achieves its 95\% CL sensitivity with only 3.09 events~\cite{ParticleDataGroup:2024cfk}.

In this section, we will explore HNLs generated from Higgs decay processes. The overall inclusive cross section for a single Higgs boson is mainly influenced by the ggF and VBF mechanisms, with $\sigma \simeq 59.1$~pb at $\sqrt{s} = 14$~TeV~\cite{ggF, VBF}. As before, we will consider the total integrated luminosity of $\mathcal{L} = 3$~ab$^{-1}$.

Given the geometry of FASER, the probability $\mathcal{P}$ of a HNL decaying inside the detector is
\begin{equation}
    \mathcal{P} = \left[e^{-(L - \Delta)/d} - e^{-L/d}\right]\, \Theta\left[R - L\, \tan\theta\right],
\end{equation}
where $d$ is the decay length in the laboratory frame of the HNL, and $\theta$ the angle between the momentum and the beam line. This decay length takes into account the Lorentz boost factor
\begin{equation}
    d = c \tau\, \beta\, \gamma = c \tau\, \frac{|\vec{p}|}{m_N}\,.
\end{equation}

%%%%%%%%%%%%%%%%%%%%%%%%%%%%%%%%%%%%%%%%%%%%%%%%%%%
\begin{figure}[t!]
    \def\sepf{0.49}
    \centering
    \includegraphics[width=\sepf\columnwidth]{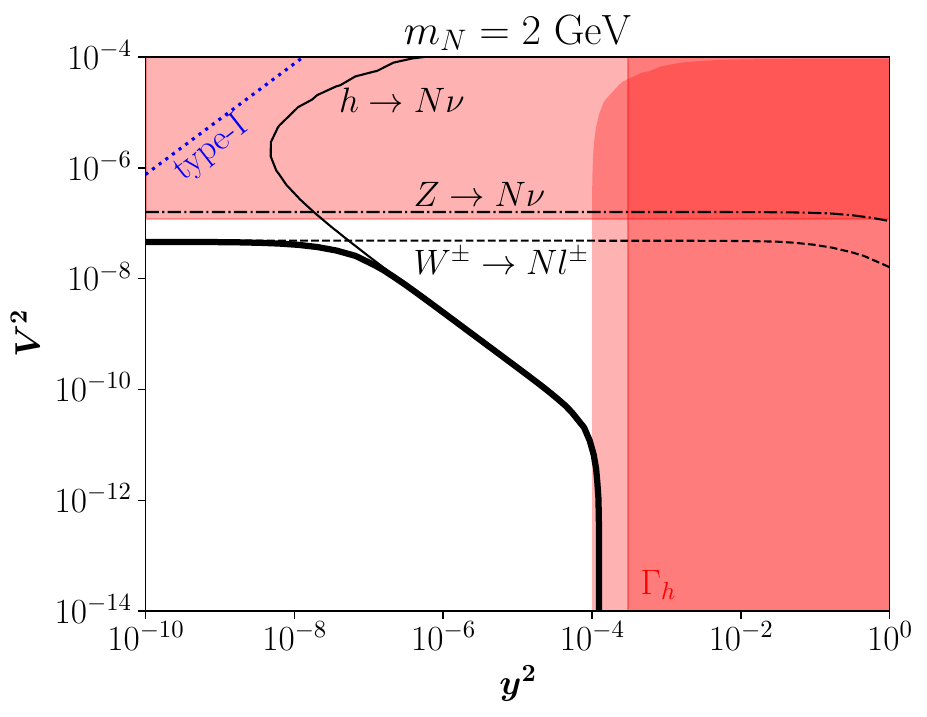}
    \includegraphics[width=\sepf\columnwidth]{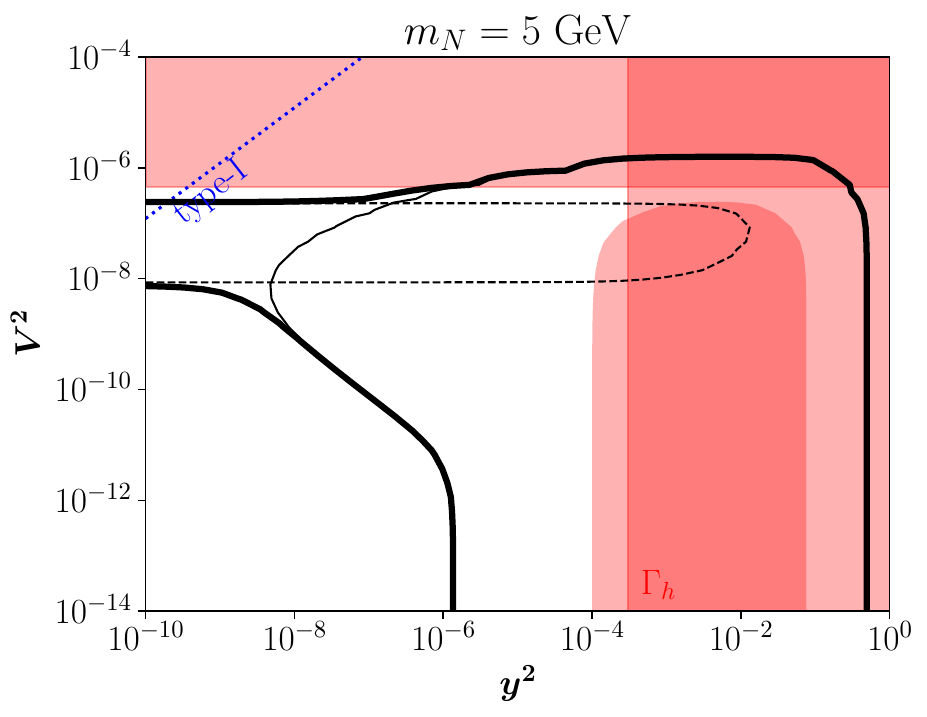}
    \includegraphics[width=\sepf\columnwidth]{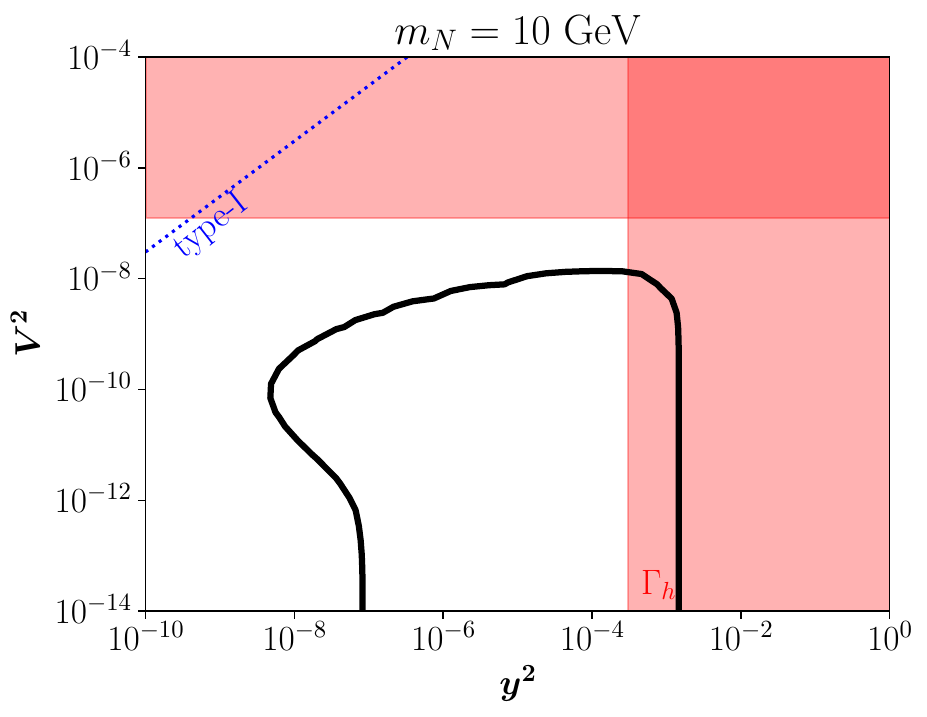}
    \includegraphics[width=\sepf\columnwidth]{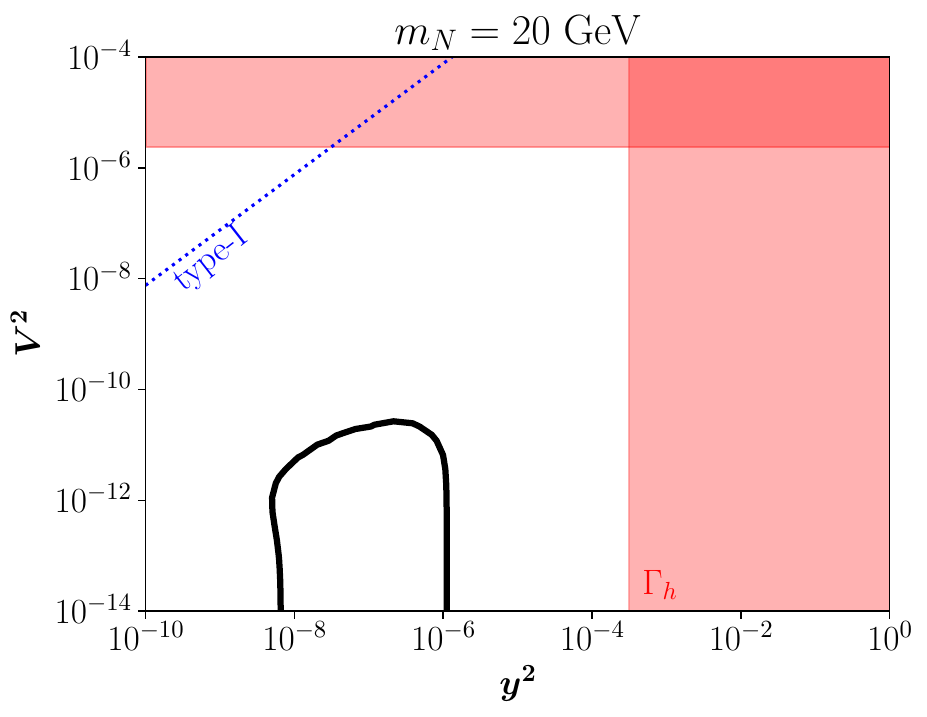}
    \caption{The sensitivity reaches for FASER are represented by the region inside the solid black lines, for $\sqrt{s} = 14$~TeV and $\mathcal{L} = 3$~ab$^{-1}$. The blue dotted line corresponds to the type-I seesaw relation. The red bands are in tension with the experimental measurements.}
    \label{fig:faser}
\end{figure} 
%%%%%%%%%%%%%%%%%%%%%%%%%%%%%%%%%%%%%%%%%% 
Figure~\ref{fig:faser} shows, with a thick solid black line, the sensitivity reach of FASER-2 to HNLs of masses 2, 5, 10 and 20~GeV in the plane $[y^2,\, |V|^2]$, where the parameter regions inside the contours will be probed. Here, the HNLs can be produced from the decays of Higgs and gauge bosons, and we allow all possible decay modes of the HNL. For completeness, the partial contributions of the different production modes are also shown with thin lines.

%%%%%%%%%%%%%%%%%%%%%%%%%%%%%%%%%%%%%%%%%%%%%%%%%%%%%%%%%%
\section{FCC-ee} \label{sec:FCCee}
%%%%%%%%%%%%%%%%%%%%%%%%%%%%%%%%%%%%%%%%%%%%%%%%%%%%%%%%%%
The initial phase of the FCC integrated project involves an $e^+e^-$ collider known as FCC-ee~\cite{Agapov:2022bhm}. This collider is designed to function as a Higgs factory, as well as an electroweak and top factory, delivering the highest luminosities across four distinct center-of-mass energy stages: the $Z$ pole, the $W^+W^-$ threshold, the $Zh$ production peak, and the $t\bar{t}$ threshold. Our focus lies on the $Z$-pole and $Zh$-production stages to evaluate our setup. The $Z$-pole stage is particularly significant due to the enormous number of $Z$ bosons expected to be produced by the end of its operation ($\sim 6 \times 10^{12}$ $Z$ bosons), providing a substantial source of HNLs, which can subsequently decay through off-shell gauge bosons and Higgs. Meanwhile, the $Zh$ production mode offers a direct mechanism to produce HNLs via Higgs decays, serving as a complementary probe to the zero-mixing scenario proposed in Ref.~\cite{Bernal:2023coo}.

%%%%%%%%%%%%%%%%%%%%%%%%%%%%%%%%%%%%%%%%%%%%%%%%%%%%%%%%%%
\subsection[$Z$ Pole]{\boldmath $Z$ Pole} \label{sec:FCCee-zpole}
%%%%%%%%%%%%%%%%%%%%%%%%%%%%%%%%%%%%%%%%%%%%%%%%%%%%%%%%%%
The $Z$-pole phase of the FCC-ee corresponds to a center-of-mass energy of $\sqrt{s} = 91.2$~GeV. This phase is characterized by exceptionally high luminosity, making it an ideal environment for precision measurements and rare process searches. During the first two years of operation, the collider is expected to achieve a luminosity of $\mathcal{L} = 34$~ab$^{-1}$ per year. This luminosity will double in the following two years, reaching $\mathcal{L} = 68$~ab$^{-1}$ per year. Therefore, the total integrated luminosity for this phase will be $\mathcal{L} = 204$~ab$^{-1}$.

The process of interest at the $Z$ pole is the production of HNLs through $Z$ decays. The relevant process chain can be expressed as\footnote{HNLs decaying to $\mu j j $ final state (through an off-shell $W$) could have a greater reach in the $V^2$ vs. $m_N$ plane. However, we do not focus on that channel as the Yukawa coupling contribution is very suppressed.}
\begin{equation}
    e^+ e^- \to Z \to \nu N \to \nu \bar{\nu} b \bar{b}\,.
\end{equation}
Here, the $Z$ boson decays into a light neutrino $\nu$ and an HNL $N$. The HNL
subsequently decays into a light neutrino and a pair of bottom quarks $b \bar{b}$ through an off-shell $Z$ or a Higgs. The main SM backgrounds contributing to the process are $e^+e^- \to b \bar b \nu \bar\nu$ with a cross section of 0.003~pb and $e^+e^- \to b \bar b$ with a much larger cross section of 8735~pb, where jet energy mismeasurements can mimic $\slashed{E}_T$.

In the next subsections, the cases where the HNL decays promptly and with a displaced vertex inside the inner tracker will be analyzed separately.

%%%%%%%%%%%%%%%%%%%%%%%%%%%%%%%%%%%%%%%%%%%%%%%%%%%%%%%%%%
\subsubsection{Prompt Decays} \label{sec:z-prompt}
%%%%%%%%%%%%%%%%%%%%%%%%%%%%%%%%%%%%%%%%%%%%%%%%%%%%%%%%%%
The optimization of the selection regions for the prompt decays includes the following cuts:
\begin{itemize}
    \item \textbf{S1:} Event must have at least two $b$-tagged jets.
    \item \textbf{S2:} The two leading $b$ jets must satisfy $p_{T_{b_1}} < 60$~GeV and $p_{T_{b_2}} < 40$~GeV.
    \item \textbf{S3:} Missing transverse energy $\slashed{E}_T$ must exceed 20~GeV.
    \item \textbf{S4:} The invariant mass of the $b$ jets must satisfy $0.2\, m_N \leq m_{b\bar{b}} \leq 1.5\, m_N$, with $\Delta R_{b \bar{b}} \leq 2.5$.
\end{itemize}

%%%%%%%%%%%%%%%%%%%%%%%%%%%%%%%%%%%%%%%%%%%%%
\begin{figure}[t!]
    \def\sepf{0.49}
    \centering
    \includegraphics[width=\sepf\columnwidth]{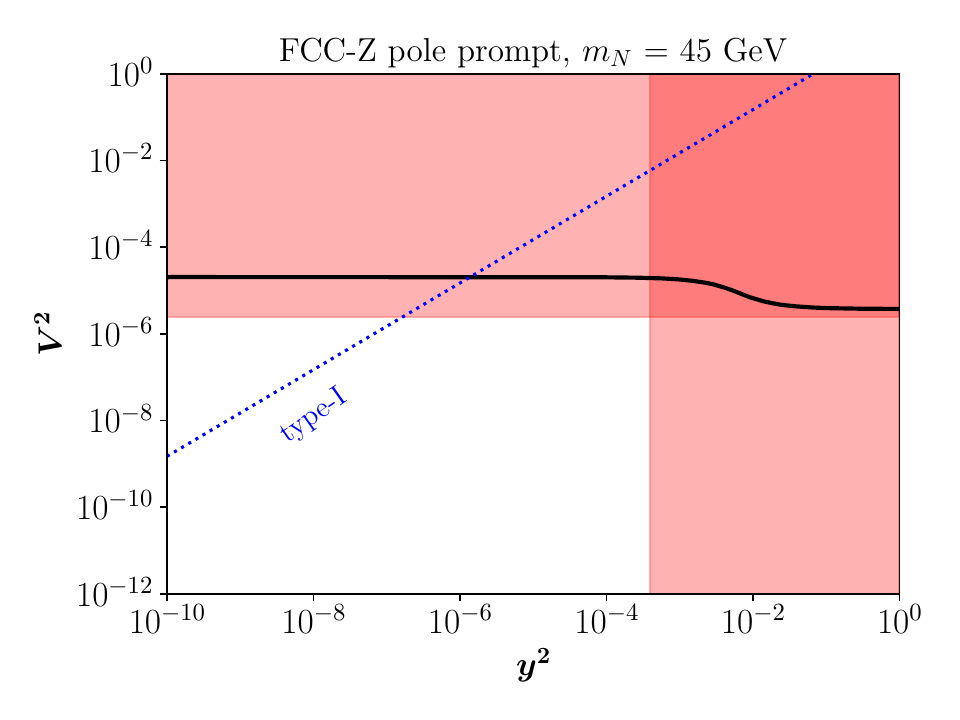}
    \includegraphics[width=\sepf\columnwidth]{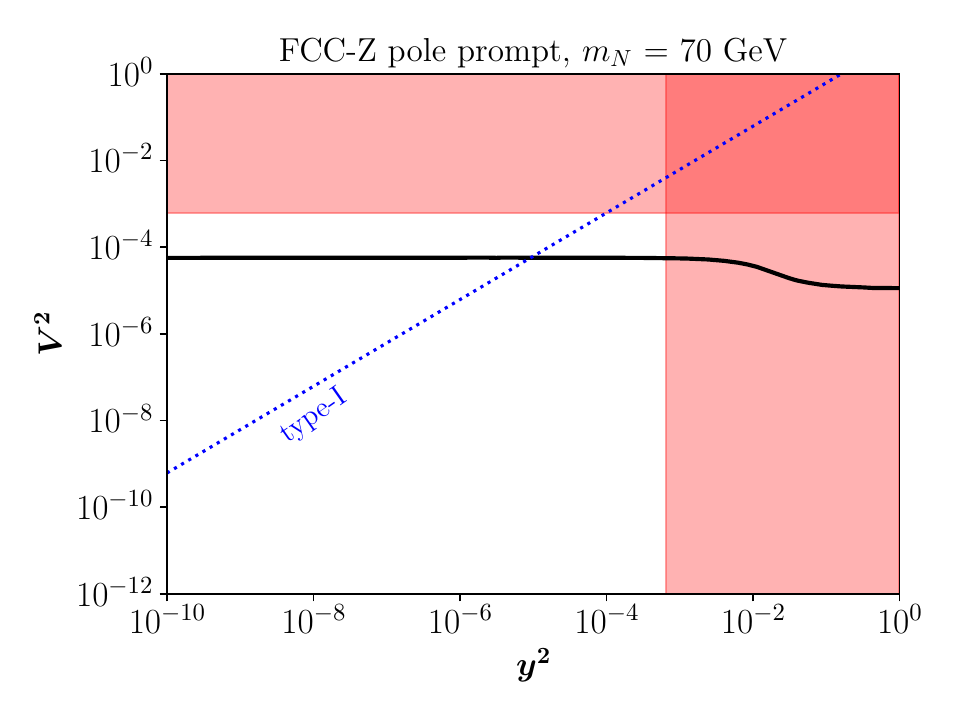}
    \caption{Discovery reach for prompt decays at FCC-ee $Z$-pole phase represented by regions above the black contours for three benchmark masses of the HNL, namely 45 and 70~GeV.}
    \label{fig:FCCeeZ-reach}
\end{figure} 
%%%%%%%%%%%%%%%%%%%%%%%%%%%%%%%%%%%%%%%%%%%%%
Figure~\ref{fig:FCCeeZ-reach} illustrates the results for prompt decays of HNLs corresponding to the benchmark mass values: $m_N = 45$ and 70~GeV. The production of HNLs requires a minimum value of the mixing parameter $V^2$, which leads to horizontal contours in the $[V^2,\,y^2]$ plane. For higher values of the Yukawa coupling $y$, the decay through the off-shell Higgs becomes dominant, as indicated by the change in the slope of the contours. 

The benchmark mass point $m_N = 45$~GeV is already excluded by current experimental constraints on the mixing parameter $V^2$. For $m_N = 70$~GeV, the parameter space is still allowed within the current bounds on the mixing. However, the region of interest, where the Yukawa coupling $y$ significantly influences the HNL decay dynamics, is already ruled out by constraints on the width of the Higgs boson $\Gamma_h$.

%%%%%%%%%%%%%%%%%%%%%%%%%%%%%%%%%%%%%%%%%%%%%%%%%%%%%%%%%%
\subsubsection{Displaced Decays} \label{sec:z-displ}
\begin{figure}[t!]
    \def\sepf{0.49}
    \centering
    \includegraphics[width=\sepf\columnwidth]{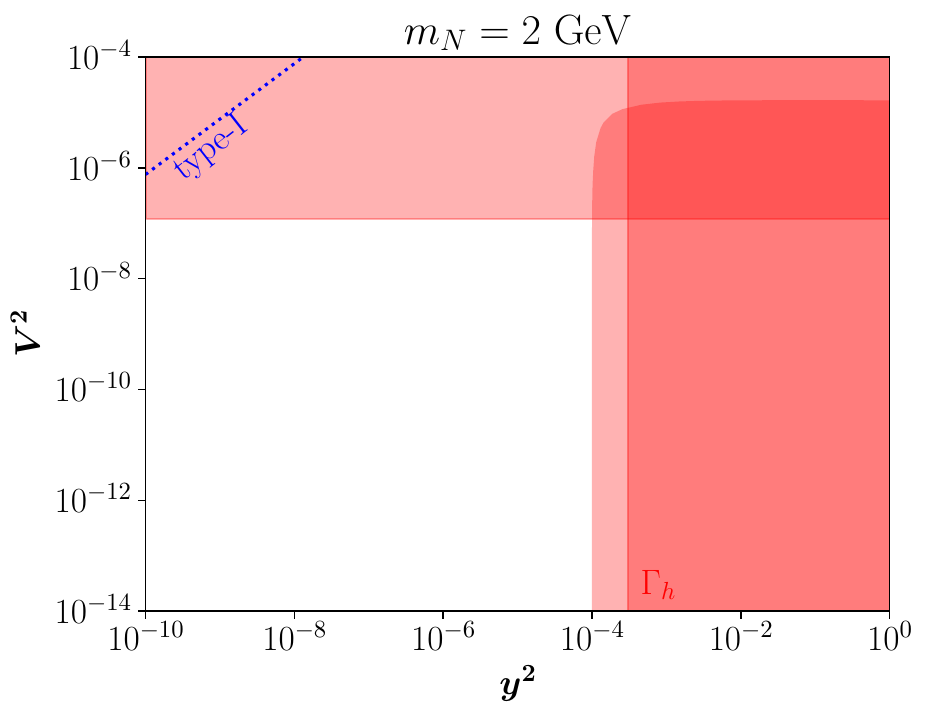}
    \includegraphics[width=\sepf\columnwidth]{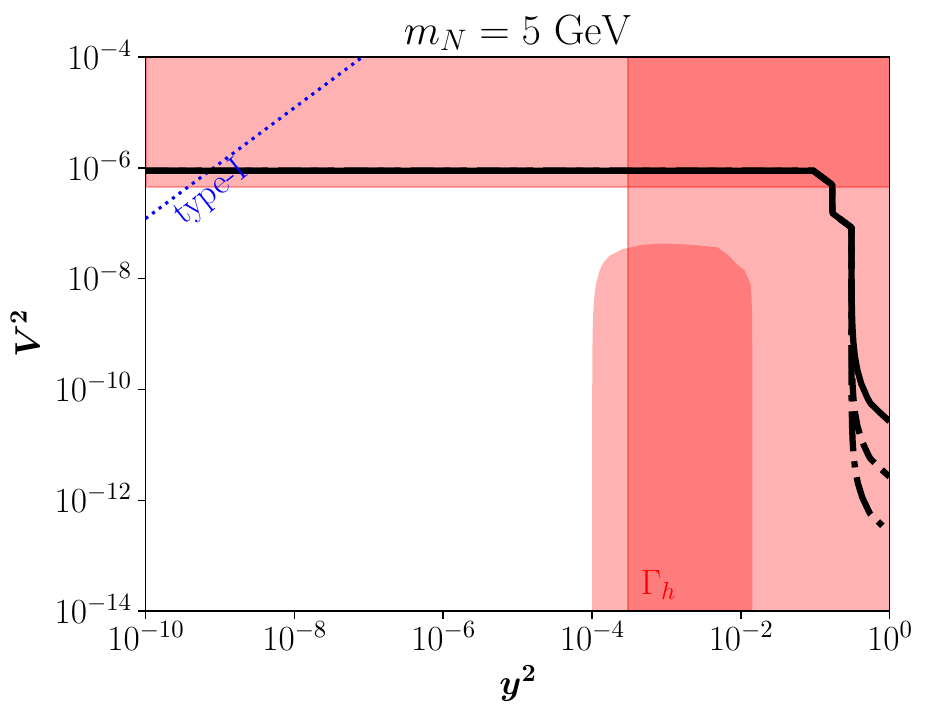}
    \includegraphics[width=\sepf\columnwidth]{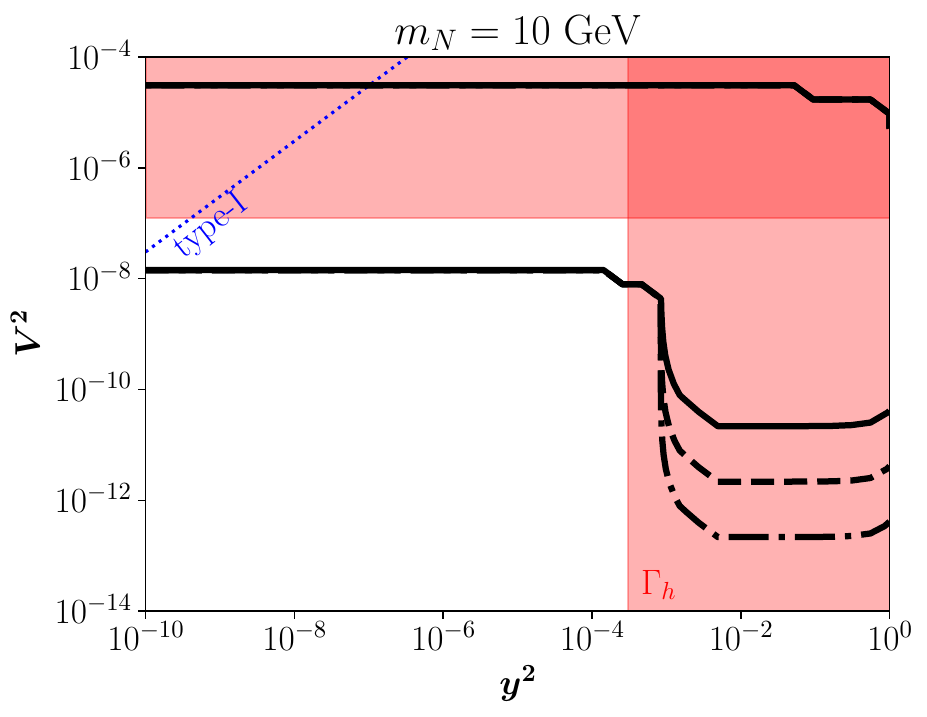}
    \includegraphics[width=\sepf\columnwidth]{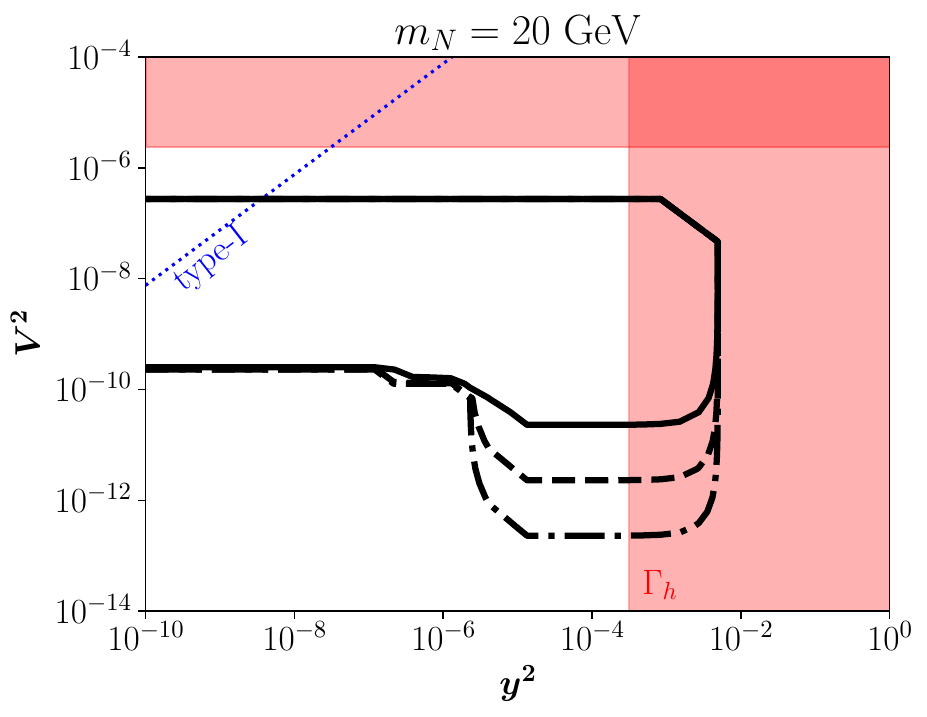}
    \caption{Sensitivity reaches for displaced vertices at FCCee in the $Z$-pole phase for all channels combined are shown by the regions inside the thick, dashed and dot-dashed contours for 100, 10 and 1 events respectively. The blue dotted line corresponds to the type-I seesaw relation. The red bands are in tension with the experimental measurements.
    }
    \label{fig:FCCeedv}
\end{figure}

%%%%%%%%%%%%%%%%%%%%%%%%%%%%%%%%%%%%%%%%%%%%%%%%%%%%%%%%%%
Figure~\ref{fig:FCCeedv} shows the results of the search for displaced vertex for HNLs with masses $m_N = 2$, 5, 10, and 20~GeV. The analysis requires two jets with transverse displacements satisfying $1~\mathrm{mm} \leq d_{xy} \leq 1$~m and longitudinal displacement $d_z \leq 300$~mm. The detection sensitivities are illustrated for 1, 10, and 100 observed events, corresponding to dot-dashed, dashed, and solid contours, respectively.
%%%%%%%%%%%%%%%%%%%%%%%%%%%%%%%%%%%%%%%%%%%%%%%%%%%
%%%%%%%%%%%%%%%%%%%%%%%%%%%%%%%%%%%%%%%%%% 

For $m_N = 2$~GeV and $m_N = 5$~GeV, the parameter space is already excluded by current experimental constraints on mixing and Higgs measurements. For $m_N = 10$~GeV, some parameter space remains accessible for mixing-sensitive searches, but the region where the Yukawa coupling $y$ plays a significant role is excluded. The most promising case is $m_N = 20$~GeV, where a substantial portion of the parameter space relevant to our setup is still allowed by the current experimental limits.

%%%%%%%%%%%%%%%%%%%%%%%%%%%%%%%%%%%%%%%%%%%%%%%%%%%%%%%%%%
\subsection[$Zh$ Production]{\boldmath $Zh$ Production} \label{sec:FCCee-zh}
%%%%%%%%%%%%%%%%%%%%%%%%%%%%%%%%%%%%%%%%%%%%%%%%%%%%%%%%%%
The $Zh$ production mode at FCC-ee is characterized by a center-of-mass energy of $\sqrt{s} = 240$~GeV, which allows for precise studies of the Higgs boson and its interactions. The planned integrated luminosity for this phase is $\mathcal{L} = 2.4$~ab$^{-1}$ per year for a period of three years, which leads to a total integrated luminosity of $\mathcal{L} = 7.2$~ab$^{-1}$.

The production cross section for the $Zh$ process is $\sigma(e^+e^- \to Zh) = 0.2403$~pb, making this mode a significant channel for Higgs studies. The associated $Z$ boson, when decaying into $e^+e^-$ or $\mu^+\mu^-$, provides a clean experimental signature for the $Zh$ process. The main background contributions come from the direct decay of $Z$ or $h$ to the charged fermion final states of the signal of which we are interested.

In the context of HNL searches, the $Zh$ production mode allows direct production of HNLs through Higgs decays, that is, $h \to N \nu$. This serves as a complementary probe to other setups, such as those based on the $Z$-pole production discussed earlier, or the zero-mixing case of HNL production through Higgs at the LHC, potentially covering regions of parameter space that are inaccessible in the $Z$-pole phase alone.

%%%%%%%%%%%%%%%%%%%%%%%%%%%%%%%%%%%%%%%%%%%%%%%%%%%%%%%%%%
\subsubsection{Without Mixing} \label{sec:fcc-ee-zh-zero-mix}
%%%%%%%%%%%%%%%%%%%%%%%%%%%%%%%%%%%%%%%%%%%%%%%%%%%%%%%%%%
For the zero-mixing case in the $Zh$ production mode at FCC-ee, the analysis focuses on prompt and displaced decays of the HNL. The prompt decay selection is based on the four criteria discussed in Section~\ref{sec:z-prompt}, combined with the requirement of at least two additional leptons arising from the decay of the associated $Z$ boson. For displaced decays, the selection criteria remain consistent with those outlined in Section~\ref{sec:z-displ}, focusing on displaced vertices with measurable transverse and longitudinal displacements.

%%%%%%%%%%%%%%%%%%%%%%%%%%%%%%%%%%%%%%%%%%%%%%%%%%%%%%%%%%
\begin{figure}[t!]
    \def\sepf{0.485}
    \centering
    \includegraphics[width=\sepf\columnwidth]{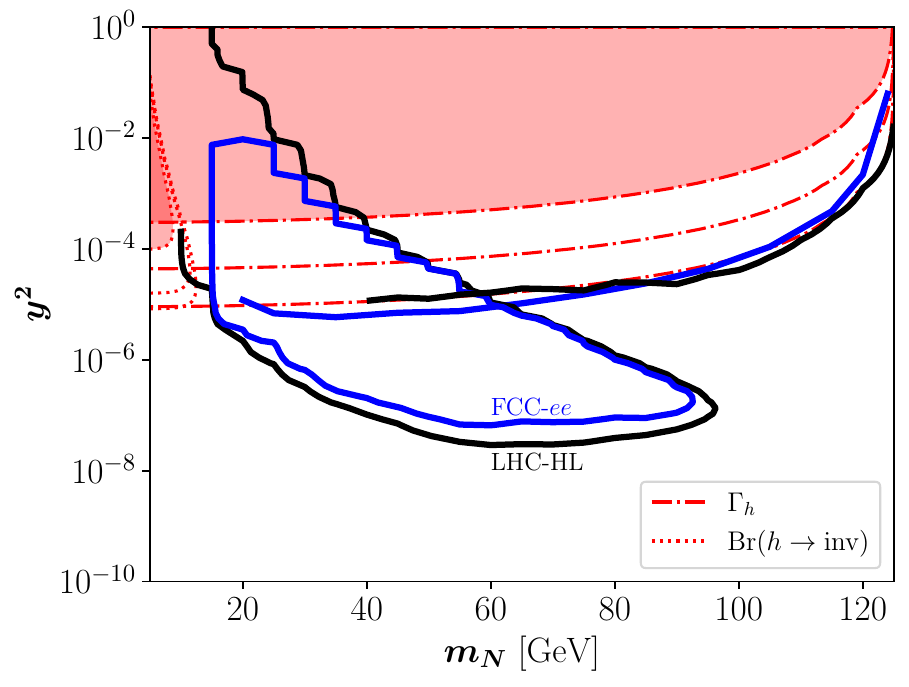}
    \includegraphics[width=0.505\columnwidth]{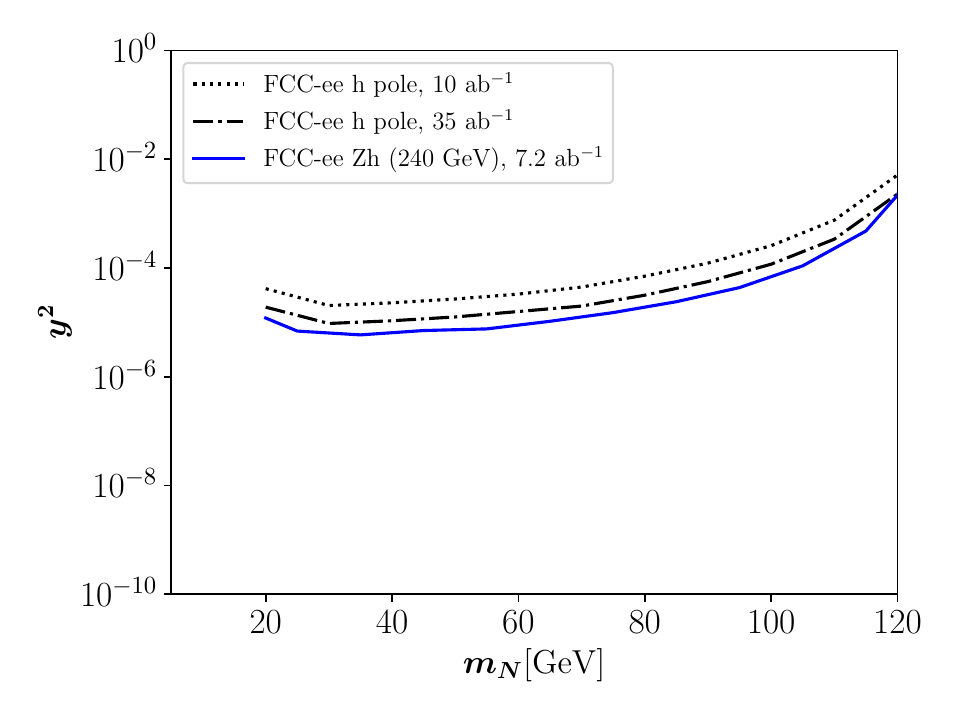}
    \caption{Left: Discovery reach comparison of Yukawa-only case between FCC-ee $Zh$ production (blue) and HL-LHC VBF Higgs production (black). Projections for the Higgs total and invisible width at HL-LHC and FCC-ee are also shown by the dot-dashed and dotted contours, respectively. Right: Comparison of the discovery reaches between HNL production at the Higgs pole and the $Zh$ prompt case, at FCC-ee.}
    \label{fig:FCCeeZH-zero-mix}
\end{figure} 
%%%%%%%%%%%%%%%%%%%%%%%%%%%%%%%%%%%%%%%%%%%%%%%%%%%%%%%%%%
The results of this analysis are illustrated in the left plot of Fig.~\ref{fig:FCCeeZH-zero-mix}. The blue contours represent the reach for both prompt and displaced decays. Compared to the sensitivity of HL-LHC for the zero-mixing case using VBF Higgs production~\cite{Bernal:2023coo} (represented by the black contours), the FCC-ee $Zh$ production mode does not provide an improvement in reach. The future projections on the invisible and total decay width of Higgs at FCC-ee, shown by the lower dotted and dot-dashed contours, can probe the prompt case completely along with a considerable amount of parameter space for the displaced case.

In the right panel of Fig.~\ref{fig:FCCeeZH-zero-mix}, the sensitivity to resonant Higgs production at its pole mass is depicted for two prospective FCC-ee operation scenarios, corresponding to luminosities of 10~ab$^{-1}$ and 35~ab$^{-1}$ (black curves). Due to the small Yukawa coupling of electrons, these sensitivities remain comparatively less significant than the one for the Higgs production through Higgstrahlung in the case of $Zh$ (blue curve).

%%%%%%%%%%%%%%%%%%%%%%%%%%%%%%%%%%%%%%%%%%%%%%%%%%%%%%%%%%
\subsubsection{With Mixing: Prompt}
%%%%%%%%%%%%%%%%%%%%%%%%%%%%%%%%%%%%%%%%%%%%%%%%%%%%%%%%%%
In the general case, the HNL can be produced from either the decay of the $Z$ boson or the Higgs boson. The analysis therefore considers both the production channels and their respective features. For HNL production through Higgs, the selection criteria remain the same as described in Section~\ref{sec:fcc-ee-zh-zero-mix}.

When the HNL is produced from the $Z$ boson decay, the Higgs decay into two leptons is considered, primarily via the $\tau^+ \tau^-$ channel. For this case, the decay of the Higgs directly into $b \bar{b}$ is deliberately avoided to maintain consistency in the final state and to circumvent the suppression arising from the efficiency of $b$-tagging, which would scale as the fourth power or more due to the presence of a minimum of four $b$ jets.

%%%%%%%%%%%%%%%%%%%%%%%%%%%%%%%%%%%%%%%%%%%%%%%%%%%%%%%%%%
\begin{figure}[t!]
    \def\sepf{0.49}
    \centering
    \includegraphics[width=\sepf\columnwidth]{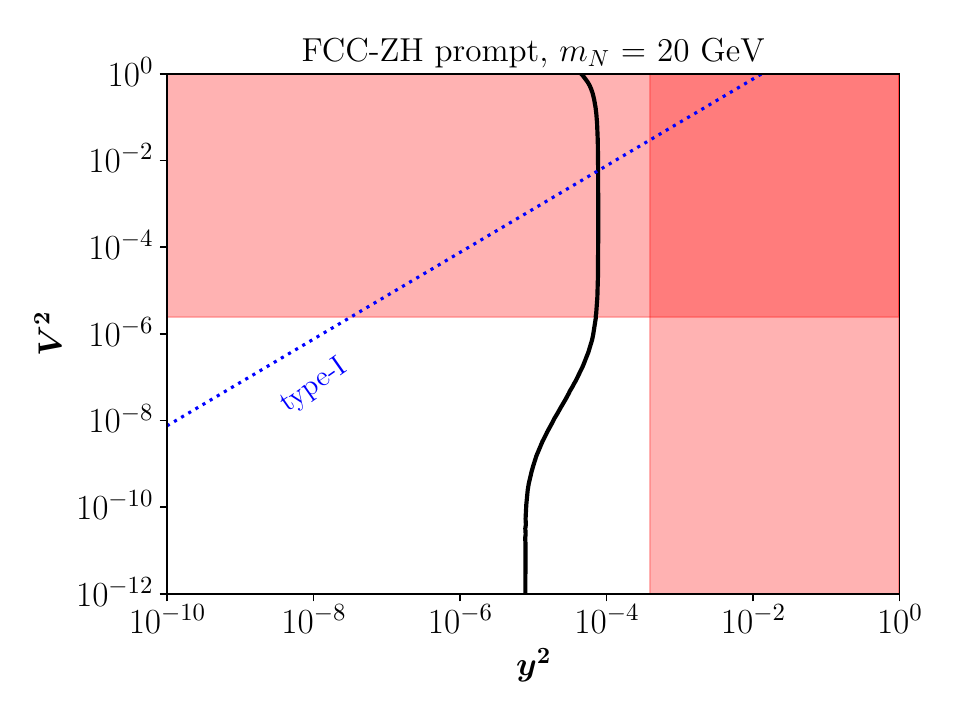}
    \includegraphics[width=\sepf\columnwidth]{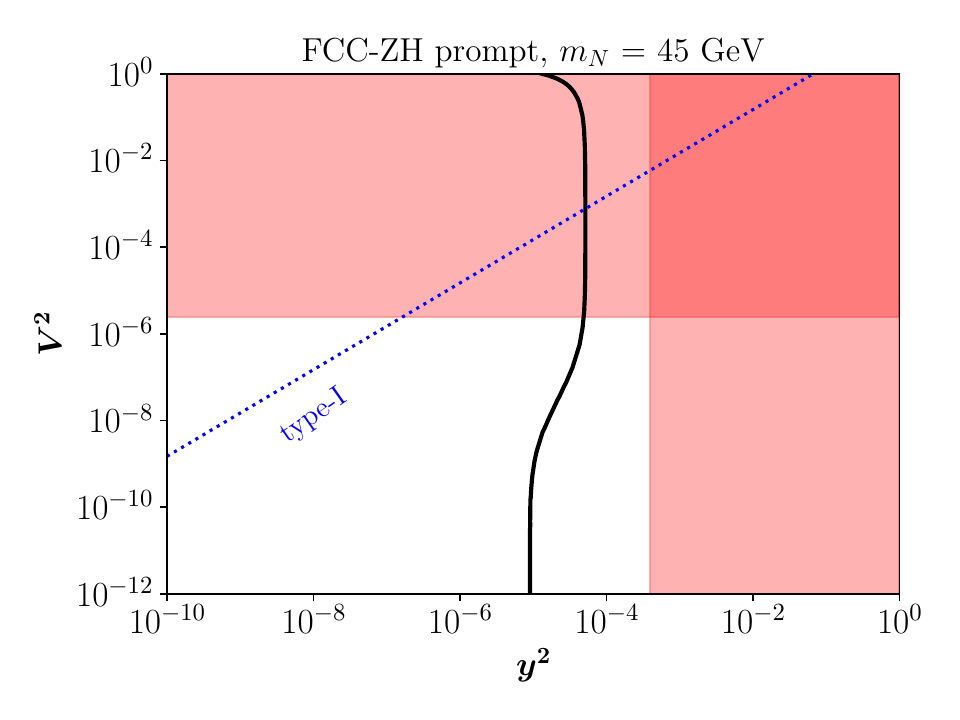}
    \includegraphics[width=\sepf\columnwidth]{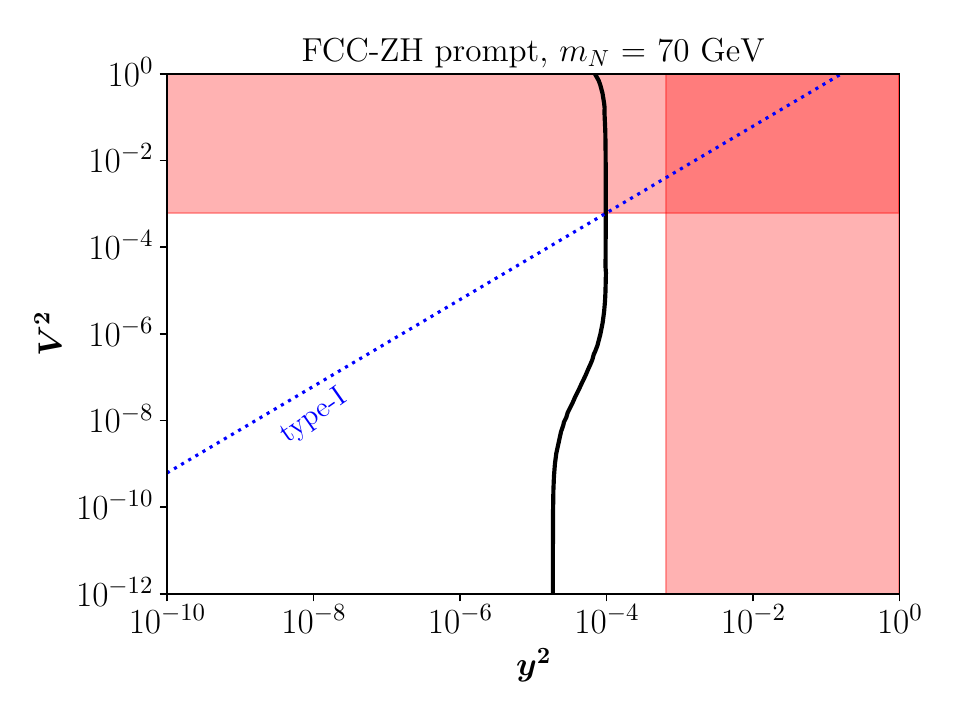}
    \caption{Discovery reach represented by regions on the right of the black contour for prompt decays at FCC-ee $Zh$ channel.}
    \label{fig:FCCeeZHprompt}
\end{figure} 
%%%%%%%%%%%%%%%%%%%%%%%%%%%%%%%%%%%%%%%%%%%%%%%%%%%%%%%%%%
The results of this analysis are presented in Fig.~\ref{fig:FCCeeZHprompt} for the HNL mass values $m_N = 20$, 45, and 70~GeV. The sensitivity is comparable to the results obtained for VBF Higgs production at the HL-LHC, as discussed in Section~\ref{sec:hl-lhc-prompt}. However, the clean experimental environment at FCC-ee provides a notable advantage, allowing a 3$\sigma$ reach even for $m_N = 20$~GeV. An additional difference with respect to HL-LHC arises at very large values of $V^2$ where HNL production from $Z$ comes into play, leading to a shift of the sensitivity contours to smaller values of $y^2$. However, it has no interesting consequences as those mixing values are already excluded by existing constraints.

%%%%%%%%%%%%%%%%%%%%%%%%%%%%%%%%%%%%%%%%%%%%%%%%%%%%%%%%%%
\subsubsection{With Mixing: Displaced}
\begin{figure}[t!]
    \def\sepf{0.49}
    \centering
    \includegraphics[width=\sepf\columnwidth]{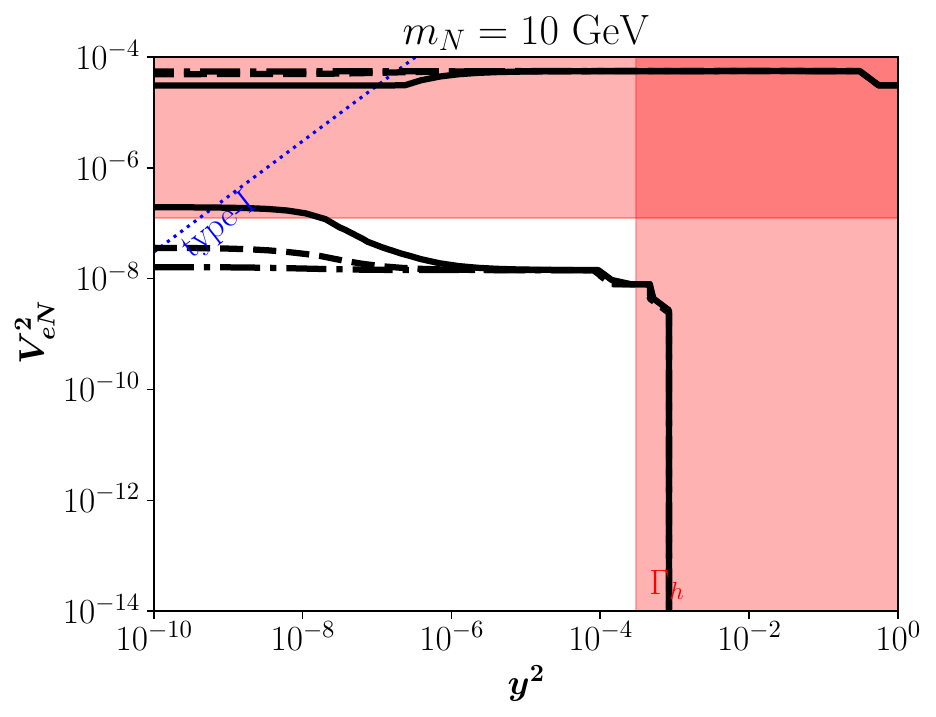}
    \includegraphics[width=\sepf\columnwidth]{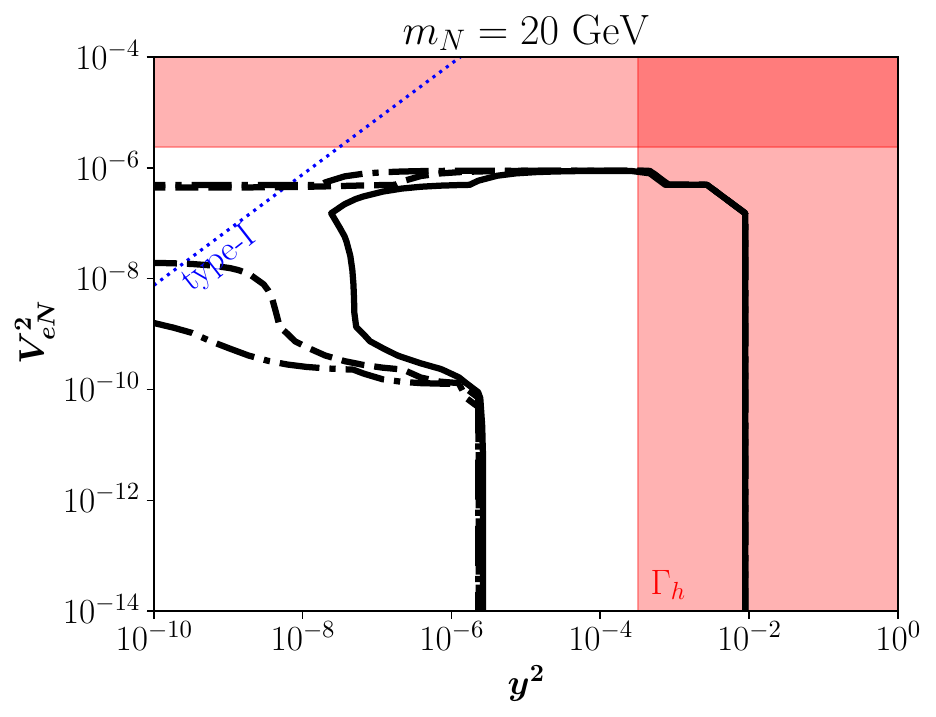}
    \includegraphics[width=\sepf\columnwidth]{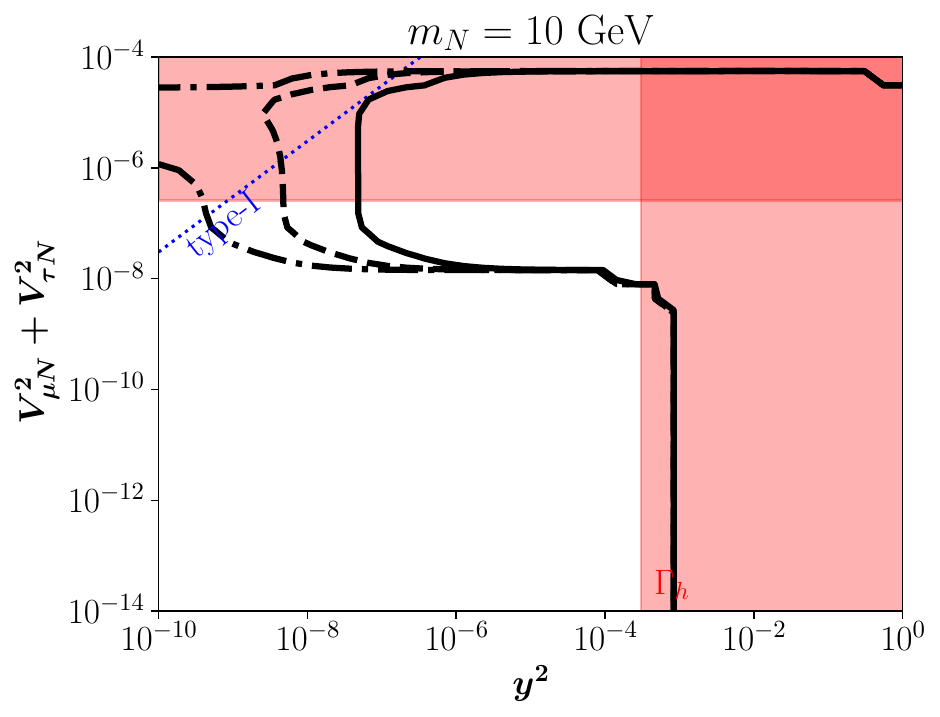}
    \includegraphics[width=\sepf\columnwidth]{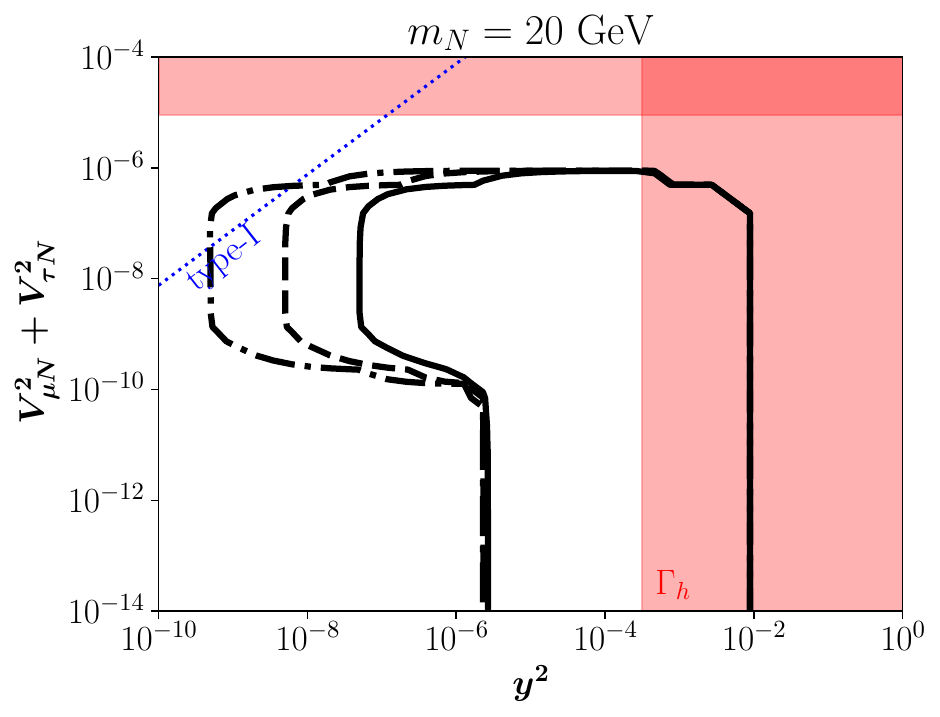}
    \caption{Sensitivity reaches for displaced vertices at FCCee with $\sqrt{s} = 240$~GeV for the $Zh$ production mode (all channels combined) are shown by regions inside the solid, dashed and dot-dashed contours of 100, 10 and 1 events. The upper two plots correspond to HNLs coupling to the electron flavor and the bottom two plots correspond to HNLs coupling to muon and tau flavors. The blue dotted line corresponds to the type-I seesaw relation. The red bands are in tension with the experimental measurements.}
    \label{fig:FCCee-ZH-dv}
\end{figure}

%%%%%%%%%%%%%%%%%%%%%%%%%%%%%%%%%%%%%%%%%%%%%%%%%%%%%%%%%%
For the displaced case, we do not apply any specific kinematic cuts to identify whether the parent particle is a $Z$ or a Higgs boson. This generalized approach allows us to consider contributions from all possible production modes of the HNL. In the case of electron flavor interactions, an additional contribution arises from the production of $N$ and $\nu_e$ through the $t$-channel $W$ exchange diagram. This additional production channel enhances the sensitivity reach for the electron-flavor case. The combined sensitivity reach, accounting for all these contributions, is illustrated in the top two panels of Fig.~\ref{fig:FCCee-ZH-dv} for $m_N$ of 10 and 20 GeV. For HNLs that interact with muon or tau flavors, the $t$-channel $W$ exchange contribution is not present. The results for these cases for the same masses are shown in the bottom two panels of Fig.~\ref{fig:FCCee-ZH-dv}.
%%%%%%%%%%%%%%%%%%%%%%%%%%%%%%%%%%%%%%%%%%%%%%%%%%%
%%%%%%%%%%%%%%%%%%%%%%%%%%%%%%%%%%%%%%%%%%%%%%%%%%%

%%%%%%%%%%%%%%%%%%%%%%%%%%%%%%%%%%%%%%%%%%%%%%%%%%%%%%%%%%
\section{Conclusions} \label{sec:concl}
%%%%%%%%%%%%%%%%%%%%%%%%%%%%%%%%%%%%%%%%%%%%%%%%%%%%%%%%%%
The extension of the Standard Model (SM) with heavy neutral leptons (HNLs) is motivated by several compelling theoretical and phenomenological considerations, such as dark matter, baryogenesis via leptogenesis, and neutrino masses. In general, HNLs can couple to SM particles through Yukawa couplings with the Higgs, and mixings with active neutrinos. In specific models, there is a tied relation between the Yukawa couplings and the mixing angles. However, one can also find scenarios where the Yukawa coupling to the SM Higgs is zero, or where mixings with active neutrinos vanish.

In this work, we remain skeptical about the possible relation between the Yukawa couplings and the mixing angles, tackling the problem without introducing mundane prejudices about the couplings of the HNLs. In order to explore the HNL parameter space, we performed a detailed collider phenomenology study within a model-independent framework, where no direct correlation is assumed between the active-sterile mixing angle and the HNL Yukawa coupling.

Our analysis considered the sensitivity of both prompt and displaced decay signatures at the HL-LHC, utilizing interaction point detectors as well as far detectors such as FASER. We also evaluated the sensitivity reach of the FCC-ee, focusing on its operation at the $Z$ pole center-of-mass energy and during the optimized $Zh$ production mode. For the scenario with no active-sterile mixing, we found no significant enhancement in sensitivity to the Yukawa coupling when comparing the HL-LHC and FCC-ee. Notably, the precision measurement of the Higgs boson width emerges as the strongest constraint on the HNL Yukawa coupling in this case. In scenarios allowing nonzero mixing, the sensitivity is dominated by the active-sterile mixing angle, which provides stringent constraints on the parameter space, especially at FCC-ee. In general, large regions of the parameter space will be probed in HL-LHC and next-generation colliders.

All in all, these results highlight the complementarity between collider experiments in probing HNL properties and emphasize the crucial role of precision Higgs measurements in constraining new physics beyond the SM.

%%%%%%%%%%%%%%%%%%%%%%%%%%%%%%%%%%%%%%%%%%%
\acknowledgments
%%%%%%%%%%%%%%%%%%%%%%%%%%%%%%%%%%%%%%%%%%%
NB received funding from Grant PID2023-151418NB-I00 funded by MCIU/AEI/10.13039/ 501100011033/ FEDER, UE.

%%%%%%%%%%%%%%%%%%%%%%%%%%%%%%%%%%%%%%%%%% 
\appendix
%%%%%%%%%%%%%%%%%%%%%%%%%%%%%%%%%%%%%%%%%% 
\section{Appendix} \label{appendix}
%%%%%%%%%%%%%%%%%%%%%%%%%%%%%%%%%%%%%%%%%% 
In order to assess whether existing experimental constraints on the mixing angle $V^2$ can impose observable limits in the general case where $y$ is treated as an independent parameter, we recast the current limits from the $[V^2,\,m_N]$ plane onto the $[V^2,\,y^2]$ plane, showing contours for fixed mass values. The recast is performed by matching the predicted number of events in our framework with the observed events at the experimental facilities. Since the factors of luminosity and efficiency cancels out on both sides, the equation for a fixed value of HNL mass reduces to
\begin{equation}
     V^2 (m_N) \times \text{BR}^{(Z^*/W^*/h^*)}_{N \rightarrow \nu f \bar{f}}(V^2, y^2, m_N) = V^2_{\text{exp}} (m_N) \times \text{BR}^{(Z^*/W^*)}_{N \rightarrow \nu f \bar{f}} (m_N)\,,
\end{equation}
where $V^2_{\text{exp}}$  and $\text{BR}^{(Z^*/W^*)}_{N \rightarrow \nu f \bar{f}}$ represents the experimental constraint and branching ratio of HNL to the final state fermions ($f$) through mixing at $m_N$. $V^2$ represents the mixing applicable to our setup at the same mass and $\text{BR}^{(Z^*/W^*/h^*)}_{N \rightarrow \nu f \bar{f}}$ is the branching ratio of HNL to the final state fermions ($f$) in our setup, which now depends on both $V^2$ and $y^2$ for a given $m_N$.

%%%%%%%%%%%%%%%%%%%%%%%%%%%%%%%%%%%%%%%%%%%%%%%%%%%%%%%%%%
\begin{figure}[t!]
    \def\sepf{0.49}
    \centering
    \includegraphics[width=\sepf\columnwidth]{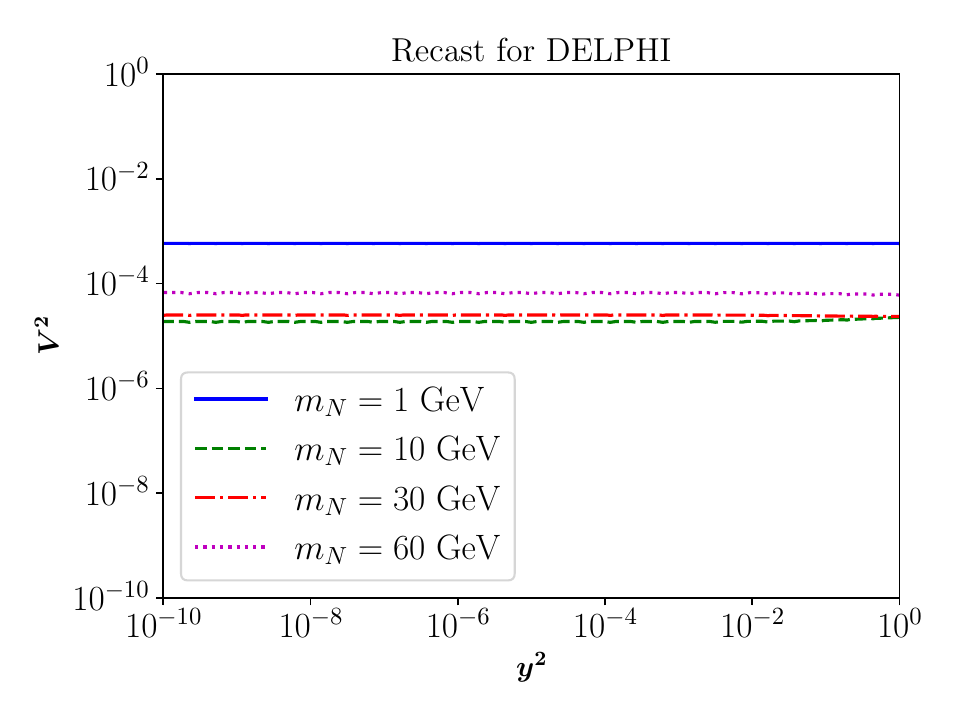}
    \includegraphics[width=\sepf\columnwidth]{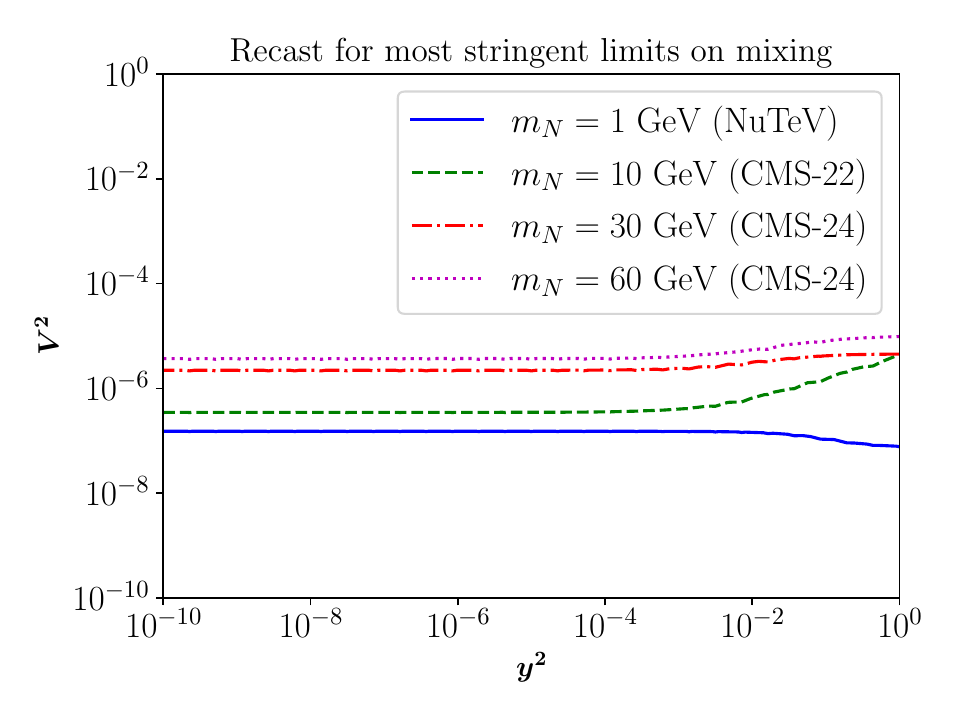}
    \caption{Recast showing contours for different values of HNL mass in $[V^2,\,y^2]$ plane. On the left is DELPHI at LEP and on the right is the most stringent experimental limit on the masses concerned.}
    \label{fig:recast}
\end{figure} 
%%%%%%%%%%%%%%%%%%%%%%%%%%%%%%%%%%%%%%%%%%%%%%%%%%%%%%%%%%
We first perform this analysis using the DELPHI constraints at LEP~\cite{Abreu:1996pa}, as its focus on jet-based final states aligns well with aspects of our setup. The results, shown in the left panel of Fig.~\ref{fig:recast}, indicate that there is no observable impact on $V^2$ with increasing $y^2$ for values $m_N$ of 1, 10, 30, and 60~GeV. 

Next, we extend this exercise by considering the most stringent existing bounds on the mixing angle for these four mass points, as presented in the right panel of Fig.~\ref{fig:recast}. In this case, the contours exhibit a change in slope at very high values of $y^2$. This is somewhat expected because the HNL production in these analyses occurs via charged- or neutral-current processes. These regions are likely already excluded by constraints from the invisible decay width and total decay width of the Higgs boson, and thus provide us with no additional insight into this general setup, underscoring the importance of the HL-LHC and FCC-ee.

%%%%%%%%%%%%%%%%%%%%%%%%%
\bibliographystyle{JHEP}
\bibliography{biblio}

\providecommand{\href}[2]{#2}\begingroup\raggedright\begin{thebibliography}{100}

\bibitem{ATLAS:2012yve}
{\scshape ATLAS} collaboration, \emph{{Observation of a new particle in the search for the Standard Model Higgs boson with the ATLAS detector at the LHC}}, \href{https://doi.org/10.1016/j.physletb.2012.08.020}{\emph{Phys. Lett. B} {\bfseries 716} (2012) 1} [\href{https://arxiv.org/abs/1207.7214}{{\ttfamily 1207.7214}}].

\bibitem{CMS:2012qbp}
{\scshape CMS} collaboration, \emph{{Observation of a New Boson at a Mass of 125 GeV with the CMS Experiment at the LHC}}, \href{https://doi.org/10.1016/j.physletb.2012.08.021}{\emph{Phys. Lett. B} {\bfseries 716} (2012) 30} [\href{https://arxiv.org/abs/1207.7235}{{\ttfamily 1207.7235}}].

\bibitem{Barber:1979yr}
D.P.~Barber et~al., \emph{{Discovery of Three Jet Events and a Test of Quantum Chromodynamics at PETRA Energies}}, \href{https://doi.org/10.1103/PhysRevLett.43.830}{\emph{Phys. Rev. Lett.} {\bfseries 43} (1979) 830}.

\bibitem{UA1:1983crd}
{\scshape UA1} collaboration, \emph{{Experimental Observation of Isolated Large Transverse Energy Electrons with Associated Missing Energy at $\sqrt{s}= 540$ GeV}}, \href{https://doi.org/10.1016/0370-2693(83)91177-2}{\emph{Phys. Lett. B} {\bfseries 122} (1983) 103}.

\bibitem{E598:1974sol}
{\scshape E598} collaboration, \emph{{Experimental Observation of a Heavy Particle $J$}}, \href{https://doi.org/10.1103/PhysRevLett.33.1404}{\emph{Phys. Rev. Lett.} {\bfseries 33} (1974) 1404}.

\bibitem{Roy:1989zd}
D.P.~Roy and S.U.~Sankar, \emph{{$B_d^0$ - $\bar{B}_d^0$ Mixing as the Evidence for the Existence of the Top Quark}}, \href{https://doi.org/10.1016/0370-2693(90)90855-Z}{\emph{Phys. Lett. B} {\bfseries 243} (1990) 296}.

\bibitem{CDF:1994juo}
{\scshape CDF} collaboration, \emph{{Evidence for top quark production in $\bar{p}p$ collisions at $\sqrt{s} = 1.8$ TeV}}, \href{https://doi.org/10.1103/PhysRevLett.73.225}{\emph{Phys. Rev. Lett.} {\bfseries 73} (1994) 225} [\href{https://arxiv.org/abs/hep-ex/9405005}{{\ttfamily hep-ex/9405005}}].

\bibitem{Abazajian:2012ys}
K.N.~Abazajian et~al., \emph{{Light Sterile Neutrinos: A White Paper}},  \href{https://arxiv.org/abs/1204.5379}{{\ttfamily 1204.5379}}.

\bibitem{Abdullahi:2022jlv}
A.M.~Abdullahi et~al., \emph{{The present and future status of heavy neutral leptons}}, \href{https://doi.org/10.1088/1361-6471/ac98f9}{\emph{J. Phys. G} {\bfseries 50} (2023) 020501} [\href{https://arxiv.org/abs/2203.08039}{{\ttfamily 2203.08039}}].

\bibitem{Minkowski:1977sc}
P.~Minkowski, \emph{{$\mu \to e\gamma$ at a Rate of One Out of $10^{9}$ Muon Decays?}}, \href{https://doi.org/10.1016/0370-2693(77)90435-X}{\emph{Phys. Lett. B} {\bfseries 67} (1977) 421}.

\bibitem{Gell-Mann:1979vob}
M.~Gell-Mann, P.~Ramond and R.~Slansky, \emph{{Complex Spinors and Unified Theories}}, {\emph{Conf. Proc. C} {\bfseries 790927} (1979) 315} [\href{https://arxiv.org/abs/1306.4669}{{\ttfamily 1306.4669}}].

\bibitem{Yanagida:1979as}
T.~Yanagida, \emph{{Horizontal gauge symmetry and masses of neutrinos}}, {\emph{Conf. Proc. C} {\bfseries 7902131} (1979) 95}.

\bibitem{Mohapatra:1979ia}
R.N.~Mohapatra and G.~Senjanovic, \emph{{Neutrino Mass and Spontaneous Parity Nonconservation}}, \href{https://doi.org/10.1103/PhysRevLett.44.912}{\emph{Phys. Rev. Lett.} {\bfseries 44} (1980) 912}.

\bibitem{Glashow:1979nm}
S.L.~Glashow, \emph{{The Future of Elementary Particle Physics}}, \href{https://doi.org/10.1007/978-1-4684-7197-7_15}{\emph{NATO Sci. Ser. B} {\bfseries 61} (1980) 687}.

\bibitem{Schechter:1980gr}
J.~Schechter and J.W.F.~Valle, \emph{{Neutrino Masses in $SU(2)\times U(1)$ Theories}}, \href{https://doi.org/10.1103/PhysRevD.22.2227}{\emph{Phys. Rev.} {\bfseries D22} (1980) 2227}.

\bibitem{Schechter:1981cv}
J.~Schechter and J.W.F.~Valle, \emph{{Neutrino Decay and Spontaneous Violation of Lepton Number}}, \href{https://doi.org/10.1103/PhysRevD.25.774}{\emph{Phys. Rev. D} {\bfseries 25} (1982) 774}.

\bibitem{Foot:1988aq}
R.~Foot, H.~Lew, X.G.~He and G.C.~Joshi, \emph{{Seesaw Neutrino Masses Induced by a Triplet of Leptons}}, \href{https://doi.org/10.1007/BF01415558}{\emph{Z. Phys. C} {\bfseries 44} (1989) 441}.

\bibitem{Ma:1998dn}
E.~Ma, \emph{{Pathways to naturally small neutrino masses}}, \href{https://doi.org/10.1103/PhysRevLett.81.1171}{\emph{Phys. Rev. Lett.} {\bfseries 81} (1998) 1171} [\href{https://arxiv.org/abs/hep-ph/9805219}{{\ttfamily hep-ph/9805219}}].

\bibitem{Choudhury:2020cpm}
D.~Choudhury, K.~Deka, T.~Mandal and S.~Sadhukhan, \emph{{Neutrino and $Z'$ phenomenology in an anomaly-free $U(1)$ extension: role of higher-dimensional operators}}, \href{https://doi.org/10.1007/JHEP06(2020)111}{\emph{JHEP} {\bfseries 06} (2020) 111} [\href{https://arxiv.org/abs/2002.02349}{{\ttfamily 2002.02349}}].

\bibitem{Fukugita:1986hr}
M.~Fukugita and T.~Yanagida, \emph{{Baryogenesis Without Grand Unification}}, \href{https://doi.org/10.1016/0370-2693(86)91126-3}{\emph{Phys. Lett. B} {\bfseries 174} (1986) 45}.

\bibitem{Akhmedov:1998qx}
E.K.~Akhmedov, V.A.~Rubakov and A.Y.~Smirnov, \emph{{Baryogenesis via neutrino oscillations}}, \href{https://doi.org/10.1103/PhysRevLett.81.1359}{\emph{Phys. Rev. Lett.} {\bfseries 81} (1998) 1359} [\href{https://arxiv.org/abs/hep-ph/9803255}{{\ttfamily hep-ph/9803255}}].

\bibitem{Asaka:2005pn}
T.~Asaka and M.~Shaposhnikov, \emph{{The $\nu$MSM, dark matter and baryon asymmetry of the universe}}, \href{https://doi.org/10.1016/j.physletb.2005.06.020}{\emph{Phys. Lett. B} {\bfseries 620} (2005) 17} [\href{https://arxiv.org/abs/hep-ph/0505013}{{\ttfamily hep-ph/0505013}}].

\bibitem{Davidson:2008bu}
S.~Davidson, E.~Nardi and Y.~Nir, \emph{{Leptogenesis}}, \href{https://doi.org/10.1016/j.physrep.2008.06.002}{\emph{Phys. Rept.} {\bfseries 466} (2008) 105} [\href{https://arxiv.org/abs/0802.2962}{{\ttfamily 0802.2962}}].

\bibitem{Hambye:2016sby}
T.~Hambye and D.~Teresi, \emph{{Higgs doublet decay as the origin of the baryon asymmetry}}, \href{https://doi.org/10.1103/PhysRevLett.117.091801}{\emph{Phys. Rev. Lett.} {\bfseries 117} (2016) 091801} [\href{https://arxiv.org/abs/1606.00017}{{\ttfamily 1606.00017}}].

\bibitem{Deka:2021koh}
K.~Deka, T.~Mandal, A.~Mukherjee and S.~Sadhukhan, \emph{{Leptogenesis in an anomaly-free U(1) extension with higher-dimensional operators}}, \href{https://doi.org/10.1016/j.nuclphysb.2023.116213}{\emph{Nucl. Phys. B} {\bfseries 991} (2023) 116213} [\href{https://arxiv.org/abs/2105.15088}{{\ttfamily 2105.15088}}].

\bibitem{Dodelson:1993je}
S.~Dodelson and L.M.~Widrow, \emph{{Sterile-neutrinos as dark matter}}, \href{https://doi.org/10.1103/PhysRevLett.72.17}{\emph{Phys. Rev. Lett.} {\bfseries 72} (1994) 17} [\href{https://arxiv.org/abs/hep-ph/9303287}{{\ttfamily hep-ph/9303287}}].

\bibitem{Shi:1998km}
X.-D.~Shi and G.M.~Fuller, \emph{{A New dark matter candidate: Nonthermal sterile neutrinos}}, \href{https://doi.org/10.1103/PhysRevLett.82.2832}{\emph{Phys. Rev. Lett.} {\bfseries 82} (1999) 2832} [\href{https://arxiv.org/abs/astro-ph/9810076}{{\ttfamily astro-ph/9810076}}].

\bibitem{Abazajian:2001nj}
K.~Abazajian, G.M.~Fuller and M.~Patel, \emph{{Sterile neutrino hot, warm, and cold dark matter}}, \href{https://doi.org/10.1103/PhysRevD.64.023501}{\emph{Phys. Rev. D} {\bfseries 64} (2001) 023501} [\href{https://arxiv.org/abs/astro-ph/0101524}{{\ttfamily astro-ph/0101524}}].

\bibitem{Cline:2020mdt}
J.M.~Cline, M.~Puel and T.~Toma, \emph{{A little theory of everything, with heavy neutral leptons}}, \href{https://doi.org/10.1007/JHEP05(2020)039}{\emph{JHEP} {\bfseries 05} (2020) 039} [\href{https://arxiv.org/abs/2001.11505}{{\ttfamily 2001.11505}}].

\bibitem{Asaka:2005an}
T.~Asaka, S.~Blanchet and M.~Shaposhnikov, \emph{{The $\nu$MSM, dark matter and neutrino masses}}, \href{https://doi.org/10.1016/j.physletb.2005.09.070}{\emph{Phys. Lett. B} {\bfseries 631} (2005) 151} [\href{https://arxiv.org/abs/hep-ph/0503065}{{\ttfamily hep-ph/0503065}}].

\bibitem{Abreu:1996pa}
{\scshape DELPHI} collaboration, \emph{{Search for neutral heavy leptons produced in Z decays}}, \href{https://doi.org/10.1007/s002880050370}{\emph{Z. Phys.} {\bfseries C74} (1997) 57}.

\bibitem{Aad:2011vj}
{\scshape ATLAS} collaboration, \emph{{Inclusive search for same-sign dilepton signatures in $pp$ collisions at $\sqrt{s}=7$ TeV with the ATLAS detector}}, \href{https://doi.org/10.1007/JHEP10(2011)107}{\emph{JHEP} {\bfseries 10} (2011) 107} [\href{https://arxiv.org/abs/1108.0366}{{\ttfamily 1108.0366}}].

\bibitem{Chatrchyan:2012fla}
{\scshape CMS} collaboration, \emph{{Search for heavy Majorana neutrinos in $\mu^{\pm}\mu^{\pm} +$ jets and $e^{\pm}e^{\pm} +$ jets events in pp collisions at $\sqrt{s} =$ 7 TeV}}, \href{https://doi.org/10.1016/j.physletb.2012.09.012}{\emph{Phys. Lett.} {\bfseries B717} (2012) 109} [\href{https://arxiv.org/abs/1207.6079}{{\ttfamily 1207.6079}}].

\bibitem{Khachatryan:2015gha}
{\scshape CMS} collaboration, \emph{{Search for heavy Majorana neutrinos in $\mu^\pm \mu^\pm+$ jets events in proton-proton collisions at $\sqrt{s}$ = 8 TeV}}, \href{https://doi.org/10.1016/j.physletb.2015.06.070}{\emph{Phys. Lett.} {\bfseries B748} (2015) 144} [\href{https://arxiv.org/abs/1501.05566}{{\ttfamily 1501.05566}}].

\bibitem{Aad:2015xaa}
{\scshape ATLAS} collaboration, \emph{{Search for heavy Majorana neutrinos with the ATLAS detector in pp collisions at $ \sqrt{s}=8 $ TeV}}, \href{https://doi.org/10.1007/JHEP07(2015)162}{\emph{JHEP} {\bfseries 07} (2015) 162} [\href{https://arxiv.org/abs/1506.06020}{{\ttfamily 1506.06020}}].

\bibitem{Khachatryan:2016olu}
{\scshape CMS} collaboration, \emph{{Search for heavy Majorana neutrinos in e$^{}$e$^{}$+ jets and e$^{}$ $\mu^{}$+ jets events in proton-proton collisions at $ \sqrt{s}=8 $ TeV}}, \href{https://doi.org/10.1007/JHEP04(2016)169}{\emph{JHEP} {\bfseries 04} (2016) 169} [\href{https://arxiv.org/abs/1603.02248}{{\ttfamily 1603.02248}}].

\bibitem{Sirunyan:2018mtv}
{\scshape CMS} collaboration, \emph{{Search for heavy neutral leptons in events with three charged leptons in proton-proton collisions at $\sqrt{s} =$ 13 TeV}}, \href{https://doi.org/10.1103/PhysRevLett.120.221801}{\emph{Phys. Rev. Lett.} {\bfseries 120} (2018) 221801} [\href{https://arxiv.org/abs/1802.02965}{{\ttfamily 1802.02965}}].

\bibitem{delAguila:2007qnc}
F.~del {\'A}guila, J.A.~Aguilar-Saavedra and R.~Pittau, \emph{{Heavy neutrino signals at large hadron colliders}}, \href{https://doi.org/10.1088/1126-6708/2007/10/047}{\emph{JHEP} {\bfseries 10} (2007) 047} [\href{https://arxiv.org/abs/hep-ph/0703261}{{\ttfamily hep-ph/0703261}}].

\bibitem{delAguila:2008cj}
F.~del {\'A}guila and J.A.~Aguilar-Saavedra, \emph{{Distinguishing seesaw models at LHC with multi-lepton signals}}, \href{https://doi.org/10.1016/j.nuclphysb.2008.12.029}{\emph{Nucl. Phys.} {\bfseries B813} (2009) 22} [\href{https://arxiv.org/abs/0808.2468}{{\ttfamily 0808.2468}}].

\bibitem{delAguila:2008hw}
F.~del {\'A}guila and J.A.~Aguilar-Saavedra, \emph{{Electroweak scale seesaw and heavy Dirac neutrino signals at LHC}}, \href{https://doi.org/10.1016/j.physletb.2009.01.010}{\emph{Phys. Lett.} {\bfseries B672} (2009) 158} [\href{https://arxiv.org/abs/0809.2096}{{\ttfamily 0809.2096}}].

\bibitem{Atre:2009rg}
A.~Atre, T.~Han, S.~Pascoli and B.~Zhang, \emph{{The Search for Heavy Majorana Neutrinos}}, \href{https://doi.org/10.1088/1126-6708/2009/05/030}{\emph{JHEP} {\bfseries 05} (2009) 030} [\href{https://arxiv.org/abs/0901.3589}{{\ttfamily 0901.3589}}].

\bibitem{BhupalDev:2012zg}
P.S.~Bhupal~Dev, R.~Franceschini and R.N.~Mohapatra, \emph{{Bounds on TeV Seesaw Models from LHC Higgs Data}}, \href{https://doi.org/10.1103/PhysRevD.86.093010}{\emph{Phys. Rev. D} {\bfseries 86} (2012) 093010} [\href{https://arxiv.org/abs/1207.2756}{{\ttfamily 1207.2756}}].

\bibitem{Dev:2013wba}
P.S.B.~Dev, A.~Pilaftsis and U.-k.~Yang, \emph{{New Production Mechanism for Heavy Neutrinos at the LHC}}, \href{https://doi.org/10.1103/PhysRevLett.112.081801}{\emph{Phys. Rev. Lett.} {\bfseries 112} (2014) 081801} [\href{https://arxiv.org/abs/1308.2209}{{\ttfamily 1308.2209}}].

\bibitem{Das:2014jxa}
A.~Das, P.S.~Bhupal~Dev and N.~Okada, \emph{{Direct bounds on electroweak scale pseudo-Dirac neutrinos from $\sqrt s=8$~TeV LHC data}}, \href{https://doi.org/10.1016/j.physletb.2014.06.058}{\emph{Phys. Lett.} {\bfseries B735} (2014) 364} [\href{https://arxiv.org/abs/1405.0177}{{\ttfamily 1405.0177}}].

\bibitem{Alva:2014gxa}
D.~Alva, T.~Han and R.~Ruiz, \emph{{Heavy Majorana neutrinos from $W\gamma$ fusion at hadron colliders}}, \href{https://doi.org/10.1007/JHEP02(2015)072}{\emph{JHEP} {\bfseries 02} (2015) 072} [\href{https://arxiv.org/abs/1411.7305}{{\ttfamily 1411.7305}}].

\bibitem{Deppisch:2015qwa}
F.F.~Deppisch, P.S.~Bhupal~Dev and A.~Pilaftsis, \emph{{Neutrinos and Collider Physics}}, \href{https://doi.org/10.1088/1367-2630/17/7/075019}{\emph{New J. Phys.} {\bfseries 17} (2015) 075019} [\href{https://arxiv.org/abs/1502.06541}{{\ttfamily 1502.06541}}].

\bibitem{Banerjee:2015gca}
S.~Banerjee, P.S.B.~Dev, A.~Ibarra, T.~Mandal and M.~Mitra, \emph{{Prospects of Heavy Neutrino Searches at Future Lepton Colliders}}, \href{https://doi.org/10.1103/PhysRevD.92.075002}{\emph{Phys. Rev.} {\bfseries D92} (2015) 075002} [\href{https://arxiv.org/abs/1503.05491}{{\ttfamily 1503.05491}}].

\bibitem{Arganda:2015ija}
E.~Arganda, M.J.~Herrero, X.~Marcano and C.~Weiland, \emph{{Exotic $\mu\tau jj$ events from heavy ISS neutrinos at the LHC}}, \href{https://doi.org/10.1016/j.physletb.2015.11.013}{\emph{Phys. Lett.} {\bfseries B752} (2016) 46} [\href{https://arxiv.org/abs/1508.05074}{{\ttfamily 1508.05074}}].

\bibitem{Das:2015toa}
A.~Das and N.~Okada, \emph{{Improved bounds on the heavy neutrino productions at the LHC}}, \href{https://doi.org/10.1103/PhysRevD.93.033003}{\emph{Phys. Rev.} {\bfseries D93} (2016) 033003} [\href{https://arxiv.org/abs/1510.04790}{{\ttfamily 1510.04790}}].

\bibitem{Degrande:2016aje}
C.~Degrande, O.~Mattelaer, R.~Ruiz and J.~Turner, \emph{{Fully-Automated Precision Predictions for Heavy Neutrino Production Mechanisms at Hadron Colliders}}, \href{https://doi.org/10.1103/PhysRevD.94.053002}{\emph{Phys. Rev.} {\bfseries D94} (2016) 053002} [\href{https://arxiv.org/abs/1602.06957}{{\ttfamily 1602.06957}}].

\bibitem{Mitra:2016kov}
M.~Mitra, R.~Ruiz, D.J.~Scott and M.~Spannowsky, \emph{{Neutrino Jets from High-Mass $W_R$ Gauge Bosons in TeV-Scale Left-Right Symmetric Models}}, \href{https://doi.org/10.1103/PhysRevD.94.095016}{\emph{Phys. Rev.} {\bfseries D94} (2016) 095016} [\href{https://arxiv.org/abs/1607.03504}{{\ttfamily 1607.03504}}].

\bibitem{Das:2017zjc}
A.~Das, P.S.B.~Dev and C.S.~Kim, \emph{{Constraining Sterile Neutrinos from Precision Higgs Data}}, \href{https://doi.org/10.1103/PhysRevD.95.115013}{\emph{Phys. Rev. D} {\bfseries 95} (2017) 115013} [\href{https://arxiv.org/abs/1704.00880}{{\ttfamily 1704.00880}}].

\bibitem{Das:2017rsu}
A.~Das, Y.~Gao and T.~Kamon, \emph{{Heavy neutrino search via semileptonic Higgs decay at the LHC}}, \href{https://doi.org/10.1140/epjc/s10052-019-6937-7}{\emph{Eur. Phys. J. C} {\bfseries 79} (2019) 424} [\href{https://arxiv.org/abs/1704.00881}{{\ttfamily 1704.00881}}].

\bibitem{Ruiz:2017yyf}
R.~Ruiz, M.~Spannowsky and P.~Waite, \emph{{Heavy neutrinos from gluon fusion}}, \href{https://doi.org/10.1103/PhysRevD.96.055042}{\emph{Phys. Rev.} {\bfseries D96} (2017) 055042} [\href{https://arxiv.org/abs/1706.02298}{{\ttfamily 1706.02298}}].

\bibitem{Cai:2017mow}
Y.~Cai, T.~Han, T.~Li and R.~Ruiz, \emph{{Lepton Number Violation: Seesaw Models and Their Collider Tests}}, \href{https://doi.org/10.3389/fphy.2018.00040}{\emph{Front.in Phys.} {\bfseries 6} (2018) 40} [\href{https://arxiv.org/abs/1711.02180}{{\ttfamily 1711.02180}}].

\bibitem{Accomando:2017qcs}
E.~Accomando, L.~Delle~Rose, S.~Moretti, E.~Olaiya and C.H.~Shepherd-Themistocleous, \emph{{Extra Higgs boson and Z' as portals to signatures of heavy neutrinos at the LHC}}, \href{https://doi.org/10.1007/JHEP02(2018)109}{\emph{JHEP} {\bfseries 02} (2018) 109} [\href{https://arxiv.org/abs/1708.03650}{{\ttfamily 1708.03650}}].

\bibitem{Drewes:2018gkc}
M.~Drewes, J.~Hajer, J.~Klaric and G.~Lanfranchi, \emph{{NA62 sensitivity to heavy neutral leptons in the low scale seesaw model}}, \href{https://doi.org/10.1007/JHEP07(2018)105}{\emph{JHEP} {\bfseries 07} (2018) 105} [\href{https://arxiv.org/abs/1801.04207}{{\ttfamily 1801.04207}}].

\bibitem{Pascoli:2018rsg}
S.~Pascoli, R.~Ruiz and C.~Weiland, \emph{{Safe Jet Vetoes}}, \href{https://doi.org/10.1016/j.physletb.2018.08.060}{\emph{Phys. Lett.} {\bfseries B786} (2018) 106} [\href{https://arxiv.org/abs/1805.09335}{{\ttfamily 1805.09335}}].

\bibitem{Bhaskar:2023xkm}
A.~Bhaskar, Y.~Chaurasia, K.~Deka, T.~Mandal, S.~Mitra and A.~Mukherjee, \emph{{Right-handed neutrino pair production via second-generation leptoquarks}}, \href{https://doi.org/10.1016/j.physletb.2023.138039}{\emph{Phys. Lett. B} {\bfseries 843} (2023) 138039} [\href{https://arxiv.org/abs/2301.11889}{{\ttfamily 2301.11889}}].

\bibitem{Gronau:1984ct}
M.~Gronau, C.N.~Leung and J.L.~Rosner, \emph{{Extending Limits on Neutral Heavy Leptons}}, \href{https://doi.org/10.1103/PhysRevD.29.2539}{\emph{Phys. Rev.} {\bfseries D29} (1984) 2539}.

\bibitem{Nemevsek:2011hz}
M.~Nemev{\v s}ek, F.~Nesti, G.~Senjanovi{\'c} and Y.~Zhang, \emph{{First Limits on Left-Right Symmetry Scale from LHC Data}}, \href{https://doi.org/10.1103/PhysRevD.83.115014}{\emph{Phys. Rev.} {\bfseries D83} (2011) 115014} [\href{https://arxiv.org/abs/1103.1627}{{\ttfamily 1103.1627}}].

\bibitem{Helo:2013esa}
J.C.~Helo, M.~Hirsch and S.~Kovalenko, \emph{{Heavy neutrino searches at the LHC with displaced vertices}}, \href{https://doi.org/10.1103/PhysRevD.89.073005, 10.1103/PhysRevD.93.099902}{\emph{Phys. Rev.} {\bfseries D89} (2014) 073005} [\href{https://arxiv.org/abs/1312.2900}{{\ttfamily 1312.2900}}].

\bibitem{Izaguirre:2015pga}
E.~Izaguirre and B.~Shuve, \emph{{Multilepton and Lepton Jet Probes of Sub-Weak-Scale Right-Handed Neutrinos}}, \href{https://doi.org/10.1103/PhysRevD.91.093010}{\emph{Phys. Rev.} {\bfseries D91} (2015) 093010} [\href{https://arxiv.org/abs/1504.02470}{{\ttfamily 1504.02470}}].

\bibitem{Dib:2015oka}
C.O.~Dib and C.S.~Kim, \emph{{Discovering sterile Neutrinos ligther than $M_W$ at the LHC}}, \href{https://doi.org/10.1103/PhysRevD.92.093009}{\emph{Phys. Rev. D} {\bfseries 92} (2015) 093009} [\href{https://arxiv.org/abs/1509.05981}{{\ttfamily 1509.05981}}].

\bibitem{Dube:2017jgo}
S.~Dube, D.~Gadkari and A.M.~Thalapillil, \emph{{Lepton-Jets and Low-Mass Sterile Neutrinos at Hadron Colliders}}, \href{https://doi.org/10.1103/PhysRevD.96.055031}{\emph{Phys. Rev.} {\bfseries D96} (2017) 055031} [\href{https://arxiv.org/abs/1707.00008}{{\ttfamily 1707.00008}}].

\bibitem{Cottin:2018kmq}
G.~Cottin, J.C.~Helo and M.~Hirsch, \emph{{Searches for light sterile neutrinos with multitrack displaced vertices}}, \href{https://doi.org/10.1103/PhysRevD.97.055025}{\emph{Phys. Rev.} {\bfseries D97} (2018) 055025} [\href{https://arxiv.org/abs/1801.02734}{{\ttfamily 1801.02734}}].

\bibitem{Cottin:2018nms}
G.~Cottin, J.C.~Helo and M.~Hirsch, \emph{{Displaced vertices as probes of sterile neutrino mixing at the LHC}}, \href{https://doi.org/10.1103/PhysRevD.98.035012}{\emph{Phys. Rev.} {\bfseries D98} (2018) 035012} [\href{https://arxiv.org/abs/1806.05191}{{\ttfamily 1806.05191}}].

\bibitem{Dib:2018iyr}
C.O.~Dib, C.S.~Kim, N.A.~Neill and X.-B.~Yuan, \emph{{Search for sterile neutrinos decaying into pions at the LHC}}, \href{https://doi.org/10.1103/PhysRevD.97.035022}{\emph{Phys. Rev.} {\bfseries D97} (2018) 035022} [\href{https://arxiv.org/abs/1801.03624}{{\ttfamily 1801.03624}}].

\bibitem{Nemevsek:2018bbt}
M.~Nemev{\v s}ek, F.~Nesti and G.~Popara, \emph{{Keung-Senjanovi{\'c} process at the LHC: From lepton number violation to displaced vertices to invisible decays}}, \href{https://doi.org/10.1103/PhysRevD.97.115018}{\emph{Phys. Rev.} {\bfseries D97} (2018) 115018} [\href{https://arxiv.org/abs/1801.05813}{{\ttfamily 1801.05813}}].

\bibitem{Abada:2018sfh}
A.~Abada, N.~Bernal, M.~Losada and X.~Marcano, \emph{{Inclusive Displaced Vertex Searches for Heavy Neutral Leptons at the LHC}}, \href{https://doi.org/10.1007/JHEP01(2019)093}{\emph{JHEP} {\bfseries 01} (2019) 093} [\href{https://arxiv.org/abs/1807.10024}{{\ttfamily 1807.10024}}].

\bibitem{Marcano:2018fto}
X.~Marcano, \emph{{Heavy Neutral Leptons and displaced vertices at LHC}},  in \emph{{53$^{rd}$ Rencontres de Moriond on Electroweak Interactions and Unified Theories}}, pp.~311--316, 2018 [\href{https://arxiv.org/abs/1808.04705}{{\ttfamily 1808.04705}}].

\bibitem{Abada:2018ulr}
A.~Abada, N.~Bernal, M.~Losada and X.~Marcano, \emph{{Searching for Heavy Neutral Leptons with Displaced Vertices at the LHC}},  in \emph{{38$^{th}$ International Symposium on Physics in Collision}}, 12, 2018 [\href{https://arxiv.org/abs/1812.01720}{{\ttfamily 1812.01720}}].

\bibitem{Dib:2019ztn}
C.O.~Dib, C.S.~Kim and S.~Tapia~Araya, \emph{{Search for light sterile neutrinos from $W^\pm$ decays at the LHC}}, \href{https://doi.org/10.1103/PhysRevD.101.035022}{\emph{Phys. Rev. D} {\bfseries 101} (2020) 035022} [\href{https://arxiv.org/abs/1903.04905}{{\ttfamily 1903.04905}}].

\bibitem{Pilaftsis:1991ug}
A.~Pilaftsis, \emph{{Radiatively induced neutrino masses and large Higgs neutrino couplings in the standard model with Majorana fields}}, \href{https://doi.org/10.1007/BF01482590}{\emph{Z. Phys. C} {\bfseries 55} (1992) 275} [\href{https://arxiv.org/abs/hep-ph/9901206}{{\ttfamily hep-ph/9901206}}].

\bibitem{Maiezza:2015lza}
A.~Maiezza, M.~Nemev{\v s}ek and F.~Nesti, \emph{{Lepton Number Violation in Higgs Decay at LHC}}, \href{https://doi.org/10.1103/PhysRevLett.115.081802}{\emph{Phys. Rev. Lett.} {\bfseries 115} (2015) 081802} [\href{https://arxiv.org/abs/1503.06834}{{\ttfamily 1503.06834}}].

\bibitem{Gago:2015vma}
A.M.~Gago, P.~Hern{\'a}ndez, J.~Jones-P{\'e}rez, M.~Losada and A.~Moreno~Brice{\~n}o, \emph{{Probing the Type I Seesaw Mechanism with Displaced Vertices at the LHC}}, \href{https://doi.org/10.1140/epjc/s10052-015-3693-1}{\emph{Eur. Phys. J.} {\bfseries C75} (2015) 470} [\href{https://arxiv.org/abs/1505.05880}{{\ttfamily 1505.05880}}].

\bibitem{Accomando:2016rpc}
E.~Accomando, L.~Delle~Rose, S.~Moretti, E.~Olaiya and C.H.~Shepherd-Themistocleous, \emph{{Novel SM-like Higgs decay into displaced heavy neutrino pairs in $U(1)'$ models}}, \href{https://doi.org/10.1007/JHEP04(2017)081}{\emph{JHEP} {\bfseries 04} (2017) 081} [\href{https://arxiv.org/abs/1612.05977}{{\ttfamily 1612.05977}}].

\bibitem{Nemevsek:2016enw}
M.~Nemev{\v s}ek, F.~Nesti and J.C.~V\'asquez, \emph{{Majorana Higgses at colliders}}, \href{https://doi.org/10.1007/JHEP04(2017)114}{\emph{JHEP} {\bfseries 04} (2017) 114} [\href{https://arxiv.org/abs/1612.06840}{{\ttfamily 1612.06840}}].

\bibitem{Caputo:2017pit}
A.~Caputo, P.~Hern\'andez, J.~L{\'o}pez-Pav{\'o}n and J.~Salvado, \emph{{The seesaw portal in testable models of neutrino masses}}, \href{https://doi.org/10.1007/JHEP06(2017)112}{\emph{JHEP} {\bfseries 06} (2017) 112} [\href{https://arxiv.org/abs/1704.08721}{{\ttfamily 1704.08721}}].

\bibitem{Deppisch:2018eth}
F.F.~Deppisch, W.~Liu and M.~Mitra, \emph{{Long-lived Heavy Neutrinos from Higgs Decays}}, \href{https://doi.org/10.1007/JHEP08(2018)181}{\emph{JHEP} {\bfseries 08} (2018) 181} [\href{https://arxiv.org/abs/1804.04075}{{\ttfamily 1804.04075}}].

\bibitem{Liu:2018wte}
J.~Liu, Z.~Liu and L.-T.~Wang, \emph{{Enhancing Long-Lived Particles Searches at the LHC with Precision Timing Information}}, \href{https://doi.org/10.1103/PhysRevLett.122.131801}{\emph{Phys. Rev. Lett.} {\bfseries 122} (2019) 131801} [\href{https://arxiv.org/abs/1805.05957}{{\ttfamily 1805.05957}}].

\bibitem{Antusch:2017hhu}
S.~Antusch, E.~Cazzato and O.~Fischer, \emph{{Sterile neutrino searches via displaced vertices at LHCb}}, \href{https://doi.org/10.1016/j.physletb.2017.09.057}{\emph{Phys. Lett.} {\bfseries B774} (2017) 114} [\href{https://arxiv.org/abs/1706.05990}{{\ttfamily 1706.05990}}].

\bibitem{Adams:2013qkq}
{\scshape LBNE} collaboration, \emph{{The Long-Baseline Neutrino Experiment: Exploring Fundamental Symmetries of the Universe}},  \href{https://arxiv.org/abs/1307.7335}{{\ttfamily 1307.7335}}.

\bibitem{Gunther:2023vmz}
J.Y.~G\"unther, J.~de~Vries, H.K.~Dreiner, Z.S.~Wang and G.~Zhou, \emph{{Long-lived neutral fermions at the DUNE near detector}}, \href{https://doi.org/10.1007/JHEP01(2024)108}{\emph{JHEP} {\bfseries 01} (2024) 108} [\href{https://arxiv.org/abs/2310.12392}{{\ttfamily 2310.12392}}].

\bibitem{Coloma:2017ppo}
P.~Coloma, P.A.N.~Machado, I.~Mart\'inez-Soler and I.M.~Shoemaker, \emph{{Double-Cascade Events from New Physics in Icecube}}, \href{https://doi.org/10.1103/PhysRevLett.119.201804}{\emph{Phys. Rev. Lett.} {\bfseries 119} (2017) 201804} [\href{https://arxiv.org/abs/1707.08573}{{\ttfamily 1707.08573}}].

\bibitem{Blondel:2014bra}
{\scshape FCC-ee study Team} collaboration, \emph{{Search for Heavy Right Handed Neutrinos at the FCC-ee}}, \href{https://doi.org/10.1016/j.nuclphysbps.2015.09.304}{\emph{Nucl. Part. Phys. Proc.} {\bfseries 273-275} (2016) 1883} [\href{https://arxiv.org/abs/1411.5230}{{\ttfamily 1411.5230}}].

\bibitem{Antusch:2016vyf}
S.~Antusch, E.~Cazzato and O.~Fischer, \emph{{Displaced vertex searches for sterile neutrinos at future lepton colliders}}, \href{https://doi.org/10.1007/JHEP12(2016)007}{\emph{JHEP} {\bfseries 12} (2016) 007} [\href{https://arxiv.org/abs/1604.02420}{{\ttfamily 1604.02420}}].

\bibitem{Alekhin:2015byh}
S.~Alekhin et~al., \emph{{A facility to Search for Hidden Particles at the CERN SPS: the SHiP physics case}}, \href{https://doi.org/10.1088/0034-4885/79/12/124201}{\emph{Rept. Prog. Phys.} {\bfseries 79} (2016) 124201} [\href{https://arxiv.org/abs/1504.04855}{{\ttfamily 1504.04855}}].

\bibitem{Anelli:2015pba}
{\scshape SHiP} collaboration, \emph{{A facility to Search for Hidden Particles (SHiP) at the CERN SPS}},  \href{https://arxiv.org/abs/1504.04956}{{\ttfamily 1504.04956}}.

\bibitem{Bonivento:2013jag}
W.~Bonivento et~al., \emph{{Proposal to Search for Heavy Neutral Leptons at the SPS}},  \href{https://arxiv.org/abs/1310.1762}{{\ttfamily 1310.1762}}.

\bibitem{CMS:2015qur}
{\scshape CMS} collaboration, \emph{{Search for heavy Majorana neutrinos in $\mu^\pm \mu^\pm+$ jets events in proton-proton collisions at $\sqrt{s}$ = 8 TeV}}, \href{https://doi.org/10.1016/j.physletb.2015.06.070}{\emph{Phys. Lett. B} {\bfseries 748} (2015) 144} [\href{https://arxiv.org/abs/1501.05566}{{\ttfamily 1501.05566}}].

\bibitem{CMS:2018iaf}
{\scshape CMS} collaboration, \emph{{Search for heavy neutral leptons in events with three charged leptons in proton-proton collisions at $\sqrt{s} =$ 13 TeV}}, \href{https://doi.org/10.1103/PhysRevLett.120.221801}{\emph{Phys. Rev. Lett.} {\bfseries 120} (2018) 221801} [\href{https://arxiv.org/abs/1802.02965}{{\ttfamily 1802.02965}}].

\bibitem{ATLAS:2019kpx}
{\scshape ATLAS} collaboration, \emph{{Search for heavy neutral leptons in decays of $W$ bosons produced in 13~TeV $pp$ collisions using prompt and displaced signatures with the ATLAS detector}}, \href{https://doi.org/10.1007/JHEP10(2019)265}{\emph{JHEP} {\bfseries 10} (2019) 265} [\href{https://arxiv.org/abs/1905.09787}{{\ttfamily 1905.09787}}].

\bibitem{CMS:2022fut}
{\scshape CMS} collaboration, \emph{{Search for long-lived heavy neutral leptons with displaced vertices in proton-proton collisions at $ \sqrt{\mathrm{s}} $ =13 TeV}}, \href{https://doi.org/10.1007/JHEP07(2022)081}{\emph{JHEP} {\bfseries 07} (2022) 081} [\href{https://arxiv.org/abs/2201.05578}{{\ttfamily 2201.05578}}].

\bibitem{ATLAS:2022atq}
{\scshape ATLAS} collaboration, \emph{{Search for Heavy Neutral Leptons in Decays of W Bosons Using a Dilepton Displaced Vertex in s=13\,\,TeV pp Collisions with the ATLAS Detector}}, \href{https://doi.org/10.1103/PhysRevLett.131.061803}{\emph{Phys. Rev. Lett.} {\bfseries 131} (2023) 061803} [\href{https://arxiv.org/abs/2204.11988}{{\ttfamily 2204.11988}}].

\bibitem{Feng:2017uoz}
J.L.~Feng, I.~Galon, F.~Kling and S.~Trojanowski, \emph{{ForwArd Search ExpeRiment at the LHC}}, \href{https://doi.org/10.1103/PhysRevD.97.035001}{\emph{Phys. Rev. D} {\bfseries 97} (2018) 035001} [\href{https://arxiv.org/abs/1708.09389}{{\ttfamily 1708.09389}}].

\bibitem{FASER:2018eoc}
{\scshape FASER} collaboration, \emph{{FASER\textquoteright{}s physics reach for long-lived particles}}, \href{https://doi.org/10.1103/PhysRevD.99.095011}{\emph{Phys. Rev. D} {\bfseries 99} (2019) 095011} [\href{https://arxiv.org/abs/1811.12522}{{\ttfamily 1811.12522}}].

\bibitem{Pinfold:2019nqj}
J.L.~Pinfold, \emph{{The MoEDAL Experiment at the LHC\textemdash{}A Progress Report}}, \href{https://doi.org/10.3390/universe5020047}{\emph{Universe} {\bfseries 5} (2019) 47}.

\bibitem{Pinfold:2019zwp}
J.L.~Pinfold, \emph{{The MoEDAL experiment: a new light on the high-energy frontier}}, \href{https://doi.org/10.1098/rsta.2019.0382}{\emph{Phil. Trans. Roy. Soc. Lond. A} {\bfseries 377} (2019) 20190382}.

\bibitem{Chou:2016lxi}
J.P.~Chou, D.~Curtin and H.J.~Lubatti, \emph{{New Detectors to Explore the Lifetime Frontier}}, \href{https://doi.org/10.1016/j.physletb.2017.01.043}{\emph{Phys. Lett. B} {\bfseries 767} (2017) 29} [\href{https://arxiv.org/abs/1606.06298}{{\ttfamily 1606.06298}}].

\bibitem{Curtin:2018mvb}
D.~Curtin et~al., \emph{{Long-Lived Particles at the Energy Frontier: The MATHUSLA Physics Case}}, \href{https://doi.org/10.1088/1361-6633/ab28d6}{\emph{Rept. Prog. Phys.} {\bfseries 82} (2019) 116201} [\href{https://arxiv.org/abs/1806.07396}{{\ttfamily 1806.07396}}].

\bibitem{MATHUSLA:2020uve}
{\scshape MATHUSLA} collaboration, \emph{{An Update to the Letter of Intent for MATHUSLA: Search for Long-Lived Particles at the HL-LHC}},  \href{https://arxiv.org/abs/2009.01693}{{\ttfamily 2009.01693}}.

\bibitem{Bauer:2019vqk}
M.~Bauer, O.~Brandt, L.~Lee and C.~Ohm, \emph{{ANUBIS: Proposal to search for long-lived neutral particles in CERN service shafts}},  \href{https://arxiv.org/abs/1909.13022}{{\ttfamily 1909.13022}}.

\bibitem{Gligorov:2017nwh}
V.V.~Gligorov, S.~Knapen, M.~Papucci and D.J.~Robinson, \emph{{Searching for Long-lived Particles: A Compact Detector for Exotics at LHCb}}, \href{https://doi.org/10.1103/PhysRevD.97.015023}{\emph{Phys. Rev. D} {\bfseries 97} (2018) 015023} [\href{https://arxiv.org/abs/1708.09395}{{\ttfamily 1708.09395}}].

\bibitem{Cerci:2021nlb}
S.~Cerci et~al., \emph{{FACET: A new long-lived particle detector in the very forward region of the CMS experiment}}, \href{https://doi.org/10.1007/JHEP06(2022)110}{\emph{JHEP} {\bfseries 06} (2022) 110} [\href{https://arxiv.org/abs/2201.00019}{{\ttfamily 2201.00019}}].

\bibitem{Caputo:2016ojx}
A.~Caputo, P.~Hern{\'a}ndez, M.~Kekic, J.~L{\'o}pez-Pav{\'o}n and J.~Salvado, \emph{{The seesaw path to leptonic CP violation}}, \href{https://doi.org/10.1140/epjc/s10052-017-4823-8}{\emph{Eur. Phys. J.} {\bfseries C77} (2017) 258} [\href{https://arxiv.org/abs/1611.05000}{{\ttfamily 1611.05000}}].

\bibitem{Jana:2018rdf}
S.~Jana, N.~Okada and D.~Raut, \emph{{Displaced vertex signature of type-I seesaw model}}, \href{https://doi.org/10.1103/PhysRevD.98.035023}{\emph{Phys. Rev.} {\bfseries D98} (2018) 035023} [\href{https://arxiv.org/abs/1804.06828}{{\ttfamily 1804.06828}}].

\bibitem{Kling:2018wct}
F.~Kling and S.~Trojanowski, \emph{{Heavy Neutral Leptons at FASER}}, \href{https://doi.org/10.1103/PhysRevD.97.095016}{\emph{Phys. Rev. D} {\bfseries 97} (2018) 095016} [\href{https://arxiv.org/abs/1801.08947}{{\ttfamily 1801.08947}}].

\bibitem{Helo:2018qej}
J.C.~Helo, M.~Hirsch and Z.S.~Wang, \emph{{Heavy neutral fermions at the high-luminosity LHC}}, \href{https://doi.org/10.1007/JHEP07(2018)056}{\emph{JHEP} {\bfseries 07} (2018) 056} [\href{https://arxiv.org/abs/1803.02212}{{\ttfamily 1803.02212}}].

\bibitem{Dercks:2018wum}
D.~Dercks, H.K.~Dreiner, M.~Hirsch and Z.S.~Wang, \emph{{Long-Lived Fermions at AL3X}}, \href{https://doi.org/10.1103/PhysRevD.99.055020}{\emph{Phys. Rev. D} {\bfseries 99} (2019) 055020} [\href{https://arxiv.org/abs/1811.01995}{{\ttfamily 1811.01995}}].

\bibitem{Deppisch:2019kvs}
F.~Deppisch, S.~Kulkarni and W.~Liu, \emph{{Heavy neutrino production via $Z'$ at the lifetime frontier}}, \href{https://doi.org/10.1103/PhysRevD.100.035005}{\emph{Phys. Rev. D} {\bfseries 100} (2019) 035005} [\href{https://arxiv.org/abs/1905.11889}{{\ttfamily 1905.11889}}].

\bibitem{Frank:2019pgk}
M.~Frank, M.~de~Montigny, P.-P.A.~Ouimet, J.~Pinfold, A.~Shaa and M.~Staelens, \emph{{Searching for Heavy Neutrinos with the MoEDAL-MAPP Detector at the LHC}}, \href{https://doi.org/10.1016/j.physletb.2020.135204}{\emph{Phys. Lett. B} {\bfseries 802} (2020) 135204} [\href{https://arxiv.org/abs/1909.05216}{{\ttfamily 1909.05216}}].

\bibitem{Jones-Perez:2019plk}
J.~Jones-P\'erez, J.~Masias and J.D.~Ruiz-\'Alvarez, \emph{{Search for Long-Lived Heavy Neutrinos at the LHC with a VBF Trigger}}, \href{https://doi.org/10.1140/epjc/s10052-020-8188-z}{\emph{Eur. Phys. J. C} {\bfseries 80} (2020) 642} [\href{https://arxiv.org/abs/1912.08206}{{\ttfamily 1912.08206}}].

\bibitem{Hirsch:2020klk}
M.~Hirsch and Z.S.~Wang, \emph{{Heavy neutral leptons at ANUBIS}}, \href{https://doi.org/10.1103/PhysRevD.101.055034}{\emph{Phys. Rev. D} {\bfseries 101} (2020) 055034} [\href{https://arxiv.org/abs/2001.04750}{{\ttfamily 2001.04750}}].

\bibitem{Li:2023dbs}
J.~Li, W.~Liu and H.~Sun, \emph{{Z' mediated right-handed neutrinos from meson decays at the FASER}}, \href{https://doi.org/10.1103/PhysRevD.109.035022}{\emph{Phys. Rev. D} {\bfseries 109} (2024) 035022} [\href{https://arxiv.org/abs/2309.05020}{{\ttfamily 2309.05020}}].

\bibitem{Bhattacherjee:2023plj}
B.~Bhattacherjee, H.K.~Dreiner, N.~Ghosh, S.~Matsumoto, R.~Sengupta and P.~Solanki, \emph{{Light long-lived particles at the FCC-hh with the proposal for a dedicated forward detector FOREHUNT and a transverse detector DELIGHT}}, \href{https://doi.org/10.1103/PhysRevD.110.015036}{\emph{Phys. Rev. D} {\bfseries 110} (2024) 015036} [\href{https://arxiv.org/abs/2306.11803}{{\ttfamily 2306.11803}}].

\bibitem{Deppisch:2023sga}
F.F.~Deppisch, S.~Kulkarni and W.~Liu, \emph{{Sterile Neutrinos at MAPP in the B-L Model}},  11, 2023 [\href{https://arxiv.org/abs/2311.01719}{{\ttfamily 2311.01719}}].

\bibitem{Feng:2024zfe}
J.L.~Feng, A.~Hewitt, F.~Kling and D.~La~Rocco, \emph{{Simulating heavy neutral leptons with general couplings at collider and fixed target experiments}}, \href{https://doi.org/10.1103/PhysRevD.110.035029}{\emph{Phys. Rev. D} {\bfseries 110} (2024) 035029} [\href{https://arxiv.org/abs/2405.07330}{{\ttfamily 2405.07330}}].

\bibitem{Agapov:2022bhm}
I.~Agapov et~al., \emph{{Future Circular Lepton Collider FCC-ee: Overview and Status}},  in \emph{{Snowmass 2021}}, 3, 2022 [\href{https://arxiv.org/abs/2203.08310}{{\ttfamily 2203.08310}}].

\bibitem{Bernal:2023coo}
N.~Bernal, K.~Deka and M.~Losada, \emph{{Discovering heavy neutral leptons with the Higgs boson}}, \href{https://doi.org/10.1103/PhysRevD.110.055011}{\emph{Phys. Rev. D} {\bfseries 110} (2024) 055011} [\href{https://arxiv.org/abs/2311.18033}{{\ttfamily 2311.18033}}].

\bibitem{Esteban:2020cvm}
I.~Esteban, M.C.~Gonz\'alez-Garc\'ia, M.~Maltoni, T.~Schwetz and A.~Zhou, \emph{{The fate of hints: updated global analysis of three-flavor neutrino oscillations}}, \href{https://doi.org/10.1007/JHEP09(2020)178}{\emph{JHEP} {\bfseries 09} (2020) 178} [\href{https://arxiv.org/abs/2007.14792}{{\ttfamily 2007.14792}}].

\bibitem{deSalas:2020pgw}
P.F.~de~Salas, D.V.~Forero, S.~Gariazzo, P.~Mart\'\i{}nez-Mirav\'e, O.~Mena, C.A.~Ternes et~al., \emph{{2020 global reassessment of the neutrino oscillation picture}}, \href{https://doi.org/10.1007/JHEP02(2021)071}{\emph{JHEP} {\bfseries 02} (2021) 071} [\href{https://arxiv.org/abs/2006.11237}{{\ttfamily 2006.11237}}].

\bibitem{KATRIN:2021uub}
{\scshape KATRIN} collaboration, \emph{{Direct neutrino-mass measurement with sub-electronvolt sensitivity}}, \href{https://doi.org/10.1038/s41567-021-01463-1}{\emph{Nature Phys.} {\bfseries 18} (2022) 160} [\href{https://arxiv.org/abs/2105.08533}{{\ttfamily 2105.08533}}].

\bibitem{Dias:2012xp}
A.G.~Dias, C.A.~de~S.~Pires, P.S.~Rodrigues~da Silva and A.~Sampieri, \emph{{A Simple Realization of the Inverse Seesaw Mechanism}}, \href{https://doi.org/10.1103/PhysRevD.86.035007}{\emph{Phys. Rev. D} {\bfseries 86} (2012) 035007} [\href{https://arxiv.org/abs/1206.2590}{{\ttfamily 1206.2590}}].

\bibitem{Fraser:2014yha}
S.~Fraser, E.~Ma and O.~Popov, \emph{{Scotogenic Inverse Seesaw Model of Neutrino Mass}}, \href{https://doi.org/10.1016/j.physletb.2014.08.069}{\emph{Phys. Lett. B} {\bfseries 737} (2014) 280} [\href{https://arxiv.org/abs/1408.4785}{{\ttfamily 1408.4785}}].

\bibitem{Ma:2014qra}
E.~Ma and R.~Srivastava, \emph{{Dirac or inverse seesaw neutrino masses with $B-L$ gauge symmetry and $S_3$ flavor symmetry}}, \href{https://doi.org/10.1016/j.physletb.2014.12.049}{\emph{Phys. Lett. B} {\bfseries 741} (2015) 217} [\href{https://arxiv.org/abs/1411.5042}{{\ttfamily 1411.5042}}].

\bibitem{CentellesChulia:2016rms}
S.~Centelles~Chuli\'a, E.~Ma, R.~Srivastava and J.W.F.~Valle, \emph{{Dirac Neutrinos and Dark Matter Stability from Lepton Quarticity}}, \href{https://doi.org/10.1016/j.physletb.2017.01.070}{\emph{Phys. Lett. B} {\bfseries 767} (2017) 209} [\href{https://arxiv.org/abs/1606.04543}{{\ttfamily 1606.04543}}].

\bibitem{Ma:2006km}
E.~Ma, \emph{{Verifiable radiative seesaw mechanism of neutrino mass and dark matter}}, \href{https://doi.org/10.1103/PhysRevD.73.077301}{\emph{Phys. Rev. D} {\bfseries 73} (2006) 077301} [\href{https://arxiv.org/abs/hep-ph/0601225}{{\ttfamily hep-ph/0601225}}].

\bibitem{Cai:2017jrq}
Y.~Cai, J.~Herrero-Garc\'\i{}a, M.A.~Schmidt, A.~Vicente and R.R.~Volkas, \emph{{From the trees to the forest: a review of radiative neutrino mass models}}, \href{https://doi.org/10.3389/fphy.2017.00063}{\emph{Front. in Phys.} {\bfseries 5} (2017) 63} [\href{https://arxiv.org/abs/1706.08524}{{\ttfamily 1706.08524}}].

\bibitem{LHCHiggsCrossSectionWorkingGroup:2016ypw}
{\scshape LHC Higgs Cross Section Working Group} collaboration, \emph{{Handbook of LHC Higgs Cross Sections: 4. Deciphering the Nature of the Higgs Sector}},  \href{https://arxiv.org/abs/1610.07922}{{\ttfamily 1610.07922}}.

\bibitem{ParticleDataGroup:2024cfk}
{\scshape Particle Data Group} collaboration, \emph{{Review of particle physics}}, \href{https://doi.org/10.1103/PhysRevD.110.030001}{\emph{Phys. Rev. D} {\bfseries 110} (2024) 030001}.

\bibitem{deBlas:2022aow}
J.~de~Blas, J.~Gu and Z.~Liu, \emph{{Higgs boson precision measurements at a 125~GeV muon collider}}, \href{https://doi.org/10.1103/PhysRevD.106.073007}{\emph{Phys. Rev. D} {\bfseries 106} (2022) 073007} [\href{https://arxiv.org/abs/2203.04324}{{\ttfamily 2203.04324}}].

\bibitem{Freitas:2019bre}
A.~Freitas et~al., \emph{{Theoretical uncertainties for electroweak and Higgs-boson precision measurements at FCC-ee}},  \href{https://arxiv.org/abs/1906.05379}{{\ttfamily 1906.05379}}.

\bibitem{Fernandez-Martinez:2023phj}
E.~Fern\'andez-Mart\'\i{}nez, M.~Gonz\'alez-L\'opez, J.~Hern\'andez-Garc\'\i{}a, M.~Hostert and J.~L\'opez-Pav\'on, \emph{{Effective portals to heavy neutral leptons}}, \href{https://doi.org/10.1007/JHEP09(2023)001}{\emph{JHEP} {\bfseries 09} (2023) 001} [\href{https://arxiv.org/abs/2304.06772}{{\ttfamily 2304.06772}}].

\bibitem{HNlimits}
\url{https://github.com/mhostert/Heavy-Neutrino-Limits}.

\bibitem{Alwall:2014hca}
J.~Alwall, R.~Frederix, S.~Frixione, V.~Hirschi, F.~Maltoni, O.~Mattelaer et~al., \emph{{The automated computation of tree-level and next-to-leading order differential cross sections, and their matching to parton shower simulations}}, \href{https://doi.org/10.1007/JHEP07(2014)079}{\emph{JHEP} {\bfseries 07} (2014) 079} [\href{https://arxiv.org/abs/1405.0301}{{\ttfamily 1405.0301}}].

\bibitem{Frederix:2018nkq}
R.~Frederix, S.~Frixione, V.~Hirschi, D.~Pagani, H.S.~Shao and M.~Zaro, \emph{{The automation of next-to-leading order electroweak calculations}}, \href{https://doi.org/10.1007/JHEP11(2021)085}{\emph{JHEP} {\bfseries 07} (2018) 185} [\href{https://arxiv.org/abs/1804.10017}{{\ttfamily 1804.10017}}].

\bibitem{Degrande:2011ua}
C.~Degrande, C.~Duhr, B.~Fuks, D.~Grellscheid, O.~Mattelaer and T.~Reiter, \emph{{UFO - The Universal FeynRules Output}}, \href{https://doi.org/10.1016/j.cpc.2012.01.022}{\emph{Comput. Phys. Commun.} {\bfseries 183} (2012) 1201} [\href{https://arxiv.org/abs/1108.2040}{{\ttfamily 1108.2040}}].

\bibitem{Alloul:2013bka}
A.~Alloul, N.D.~Christensen, C.~Degrande, C.~Duhr and B.~Fuks, \emph{{FeynRules 2.0 - A complete toolbox for tree-level phenomenology}}, \href{https://doi.org/10.1016/j.cpc.2014.04.012}{\emph{Comput. Phys. Commun.} {\bfseries 185} (2014) 2250} [\href{https://arxiv.org/abs/1310.1921}{{\ttfamily 1310.1921}}].

\bibitem{ATLAS:2023tkt}
{\scshape ATLAS} collaboration, \emph{{Combination of searches for invisible decays of the Higgs boson using 139~fb$^{-1}$ of proton-proton collision data at $\sqrt{s}$=13 TeV collected with the ATLAS experiment}}, \href{https://doi.org/10.1016/j.physletb.2023.137963}{\emph{Phys. Lett. B} {\bfseries 842} (2023) 137963} [\href{https://arxiv.org/abs/2301.10731}{{\ttfamily 2301.10731}}].

\bibitem{CMS:2023sdw}
{\scshape CMS} collaboration, \emph{{A search for decays of the Higgs boson to invisible particles in events with a $t \bar t$ quark pair or a vector boson in proton-proton collisions at $\sqrt{s} = 13\,\text {Te}\hspace{-.08em}\text {V} $}}, \href{https://doi.org/10.1140/epjc/s10052-023-11952-7}{\emph{Eur. Phys. J. C} {\bfseries 83} (2023) 933} [\href{https://arxiv.org/abs/2303.01214}{{\ttfamily 2303.01214}}].

\bibitem{Dawson:2022zbb}
S.~Dawson et~al., \emph{{Report of the Topical Group on Higgs Physics for Snowmass 2021: The Case for Precision Higgs Physics}},  in \emph{{Snowmass 2021}}, 9, 2022 [\href{https://arxiv.org/abs/2209.07510}{{\ttfamily 2209.07510}}].

\bibitem{ATLAS:2023ynf}
{\scshape ATLAS} collaboration, \emph{{Measurement of the Z boson invisible width at $\sqrt{s}=13$ TeV with the ATLAS detector}}, \href{https://doi.org/10.1016/j.physletb.2024.138705}{\emph{Phys. Lett. B} {\bfseries 854} (2024) 138705} [\href{https://arxiv.org/abs/2312.02789}{{\ttfamily 2312.02789}}].

\bibitem{ggF}
\url{https://twiki.cern.ch/twiki/bin/view/LHCPhysics/LHCHWGGGF\_RUN2}.

\bibitem{VBF}
\url{https://twiki.cern.ch/twiki/bin/view/LHCPhysics/LHCHWGVBF}.

\bibitem{Muselli:2015kba}
C.~Muselli, M.~Bonvini, S.~Forte, S.~Marzani and G.~Ridolfi, \emph{{Top Quark Pair Production beyond NNLO}}, \href{https://doi.org/10.1007/JHEP08(2015)076}{\emph{JHEP} {\bfseries 08} (2015) 076} [\href{https://arxiv.org/abs/1505.02006}{{\ttfamily 1505.02006}}].

\bibitem{ATLAS:2011ac}
{\scshape ATLAS} collaboration, \emph{{Measurement of the inclusive and dijet cross-sections of $b$-jets in $pp$ collisions at $\sqrt{s}=7$ TeV with the ATLAS detector}}, \href{https://doi.org/10.1140/epjc/s10052-011-1846-4}{\emph{Eur. Phys. J. C} {\bfseries 71} (2011) 1846} [\href{https://arxiv.org/abs/1109.6833}{{\ttfamily 1109.6833}}].

\bibitem{ATLAS:2022zhj}
{\scshape ATLAS} collaboration, \emph{{Search for neutral long-lived particles in $pp$ collisions at $ \sqrt{s} $ = 13 TeV that decay into displaced hadronic jets in the ATLAS calorimeter}}, \href{https://doi.org/10.1007/JHEP06(2022)005}{\emph{JHEP} {\bfseries 06} (2022) 005} [\href{https://arxiv.org/abs/2203.01009}{{\ttfamily 2203.01009}}].

\bibitem{ATLAS:2023oti}
{\scshape ATLAS} collaboration, \emph{{Search for long-lived, massive particles in events with displaced vertices and multiple jets in pp collisions at $ \sqrt{s} $ = 13 TeV with the ATLAS detector}}, \href{https://doi.org/10.1007/JHEP06(2023)200}{\emph{JHEP} {\bfseries 2306} (2023) 200} [\href{https://arxiv.org/abs/2301.13866}{{\ttfamily 2301.13866}}].

\end{thebibliography}\endgroup
%%%%%%%%%%%%%%%%%%%%%%%%%

\end{document}